\pdfoutput=1
\documentclass[superscriptaddress,prd,twocolumn,nofootinbib]{revtex4-2}
\usepackage[utf8]{inputenc}
\usepackage[colorlinks=true, pdfstartview=FitV, linkcolor=red, citecolor=blue,
urlcolor=blue]{hyperref}
\usepackage[dvipdfmx]{graphicx}
\usepackage{amsmath,amssymb}
\usepackage{color}
\usepackage{physics}
\usepackage{slashed}
\usepackage{bm}

\newcommand{\bxt}{{\bm x}_\perp}
\newcommand{\bk}{{\bm k}}
\newcommand{\bkt}{{\bm k}_\perp}

\newcommand{\Nc}{N_{\text{c}}}

\newcommand{\ud}{\mathrm{d}}

\newcommand{\ue}{\mathrm{e}}

\newcommand{\GBs}{G_{\mathrm{B\,s}}}
\newcommand{\Lv}{L_{\mathrm{v}}}

\newcommand{\Det}{\mathrm{Det}}

\newcommand{\bphi}{{\bm\phi}}
\newcommand{\balpha}{{\bm\alpha}}

\newcommand{\su}{\mathfrak{su}}

\newcommand{\inv}[1]{\frac{1}{#1}}
\newcommand{\no}{\notag}

\newcommand{\tia}{\tilde{a}}

\begin{document}

\title{Polyakov-loop potential of accelerated gluonic matter and thermodynamic subtleties}

\author{Hao-Lei Chen}
\email{hlchen@shu.edu.cn}
\affiliation{Department of Physics, Shanghai University, Shanghai 200444, China}
\affiliation{Shanghai Research Center for Theoretical Nuclear Physics, National Natural Science Foundation of China and Fudan University, Shanghai 200438, China}

\author{Kenji Fukushima}
\email{fuku@nt.phys.s.u-tokyo.ac.jp}
\affiliation{Department of Physics, The University of Tokyo,
 7-3-1 Hongo, Bunkyo-ku, Tokyo 113-0033, Japan}
 
\author{Yu-Han Gao}
\affiliation{Physics Department and Center for Particle Physics and Field Theory, Fudan University, Shanghai 200438, China}

\author{Xu-Guang Huang}
\email{huangxuguang@fudan.edu.cn}
\affiliation{Physics Department and Center for Particle Physics and Field Theory, Fudan University, Shanghai 200438, China}
\affiliation{Key Laboratory of Nuclear Physics and Ion-beam Application (MOE), Fudan University, Shanghai 200433, China}
\affiliation{Shanghai Research Center for Theoretical Nuclear Physics, National Natural Science Foundation of China and Fudan University, Shanghai 200438, China}

\author{Yusuke Shimada}
\email{yusuke.shimada@yukawa.kyoto-u.ac.jp}
\affiliation{Yukawa Institute for Theoretical Physics, Kyoto University,
Kitashirakawa Oiwakecho, Sakyo-ku, Kyoto 606-8502, Japan}

\author{Zhi-Bin Zhu}
\affiliation{Physics Department and Center for Particle Physics and Field Theory, Fudan University, Shanghai 200438, China}

\begin{abstract}
  We study the one-loop Polyakov-loop effective potential in pure gluonic matter under constant acceleration. We perform the computation in both the Euclidean Rindler spacetime and the optical spacetime, which are related via a conformal transformation. The results from the two formulations correspond to physically different observables, and we clarify their connection to specific components of the energy-momentum tensor. This identification resolves a discrepancy previously noted for fields on conical backgrounds. For the Polyakov-loop expectation value, we should minimize the effective potential computed in the optical metric formulation, which concludes that real acceleration strengthens deconfining properties. We also discuss analytic continuation from real to imaginary acceleration and find a perturbatively confined phase. We point out some suggestive similarities and differences between systems under imaginary acceleration and imaginary rotation.
\end{abstract}

\maketitle

\section{Introduction}

Quantum chromodynamics (QCD) exhibits a crossover (i.e., continuous transition) from a confined hadronic gas to a deconfined quark-gluon plasma at finite temperature. In the heavy-quark limit, dynamical quarks decouple and QCD reduces to a
pure gluonic system, where this crossover becomes a genuine phase
transition. In this limit, the expectation value of the Polyakov loop,
\(L\), serves as a well-defined order parameter.  The seminal works by Gross, Pisarski, and Yaffe~\cite{Gross:1980br} and Weiss~\cite{Weiss:1980rj,Weiss:1981ev} have shown that the one-loop Polyakov-loop effective potential successfully captures expected properties of deconfined matter in the perturbative regime at sufficiently high temperature. Since then, the Polyakov-loop effective potential has become a standard tool~\cite{Fukushima:2017csk} for studying how extreme environments, such as high temperature, large baryon density,  strong electromagnetic fields, fast rotation, etc., influence the confinement-deconfinement transition in a pure gluonic system described by the Yang--Mills theory.

Among various extreme environments, acceleration remains one of the
least understood, although it is conceptually simple and
phenomenologically important in, e.g., heavy-ion collisions where the early-stage quark-gluon matter may experience large proper
acceleration~\cite{Kharzeev:2005iz,Zhong:2026ugc,Prokhorov:2025vak}. By the
equivalence principle, uniform acceleration is closely related to a
gravitational field, so acceleration provides a simple setting for studying QCD matter in non-inertial or curved geometry. Moreover, acceleration is generated by the boost operator, while rotation is generated by the angular-momentum operator. This analogy naturally motivates the question of whether acceleration plays a role similar to rotation in the deconfinement of QCD matter.

One defining feature of acceleration is the Unruh effect: an accelerated observer sees the Minkowski vacuum as a thermal bath at temperature (called the Unruh temperature) $T_{\text{U}}=a/(2\pi)$ with $a$ the proper acceleration~\cite{Fulling:1972md,Davies:1974th,Unruh:1976db,Unruh:1983ac,Crispino:2007eb}. This suggests that acceleration acts as a thermal reservoir, melts condensates, and restores broken symmetries. For example, chiral symmetry restoration has been discussed in the Nambu--Jona-Lasinio (NJL) model and related effective theories~\cite{Ohsaku:2004rv,Ebert:2006bh,Dobado:2017dvs,Dobado:2017xxb,Casado-Turrion:2019xta,Casado-Turrion:2019gbg,Kou:2024dml}. Related phenomena, such as dissociation of mesons and melting of Bose--Einstein condensates by acceleration, have also been examined~\cite{Peeters:2007ti,Paredes:2008cr,Ghoroku:2010sp,Castorina:2012yg,Takeuchi:2015nga}. In most of these analyses~\cite{Ohsaku:2004rv,Ebert:2006bh,Dobado:2017dvs,Dobado:2017xxb,Casado-Turrion:2019xta,Casado-Turrion:2019gbg,Castorina:2012yg,Takeuchi:2015nga}, the system is taken to be in a thermal Rindler state at temperature $T=T_{\text{U}}$, corresponding to a Minkowski vacuum. The critical acceleration for symmetry restoration is found to be $a_c=2\pi T_c$ with $T_c$ the critical temperature in Minkowski spacetime. This result agrees with the intuitive expectation from the Unruh effect. 

This intuitive picture, however, is not yet complete. The phase structure of accelerated systems depends sensitively on how the vacuum contribution is subtracted~\cite{Benic:2015qha,Salluce:2024jlj}. It was found that subtracting the Rindler vacuum leaves the critical temperature of chiral phase transition unchanged, whereas subtracting the Minkowski vacuum makes it grow with acceleration~\cite{Chernodub:2025ovo,Zhu:2025pxh}. A first-principles lattice study of weakly accelerated gluonic matter has also been performed, but the acceleration dependence of the deconfinement temperature remains to be further clarified~\cite{Chernodub:2024wis,Braguta:2026nfy}. Therefore, it is a timely subject to compute the Polyakov-loop effective potential directly and reveal the phase-transition properties of a hot and accelerated gluonic system.

We should emphasize that the field-theoretical formulation in Rindler
spacetime at a general temperature is subtle.  This is because the Euclidean continuation from Rindler spacetime develops a conical singularity whenever $T\neq T_{\text{U}}$; see explanations in Sec.~\ref{sec:euc-opt}. This situation is closely related to quantum field theory on conical manifolds and cosmic-string backgrounds~\cite{Linet:1987vz,Fursaev:1993qk,Dowker:1994fi,Solodukhin:1994yz,Fursaev:1995ef,Fursaev:1996uz}. For spin-one fields, surface or contact terms localized at the conical tip are known to affect thermodynamic quantities~\cite{Kabat:1995eq,Iellici:1996jx,Moretti:1996ws}. In the commonly used formulation, the Rindler metric is conformally mapped to an ultrastatic optical metric~\cite{Gibbons:1976pt,Dowker:1978md,Gusev:1998rp,Emparan:1994qa,Bytsenko:1995ym,Cognola:1997dv,Frolov:1998vs,Sonego:2010vy} where finite-temperature calculations become more tractable. We closely discuss these formulations in Secs.~\ref{sec:gene-two-met} and \ref{sec:poly-pot}.

We thus employ two complementary approaches to study the Polyakov-loop effective potential in accelerated gluonic matter. First, we work directly in the Rindler spacetime. Using the Euclidean path-integral formalism, we compute both the
local form of the partition function and the energy-momentum tensor (EMT). Second, we work in the optical spacetime, where acceleration is encoded in the curvature. In this latter setting, we employ the heat-kernel expansion to construct the Polyakov-loop effective potential. 

These would-be equivalent approaches yield different results for the Polyakov-loop effective potential. Similarly controversial results have been reported for scalars, spinors~\cite{Fursaev:1997th,Diakonov:2023jdk}, and vectors~\cite{Moretti:1996wd}. A related subtlety arises also in the density-operator approach to global equilibrium of accelerated fermions and bosons~\cite{Zubarev:1979afm,Prokhorov:2019cik,Prokhorov:2019yft,Becattini:2020qol,Palermo:2021hlf,Palermo:2023ews,Becattini:2025oyi}. In a recent work~\cite{Ambrus:2025dca}, it has been confirmed that the pressure derived from the EMT could differ from the pressure obtained from the partition function.

Our calculations explicitly demonstrate that the Polyakov-loop effective potentials obtained in these two approaches are connected by a thermodynamic relation, once we correctly identify the results as the EMT components.  More specifically, the effective potential obtained directly in the Rindler
description is somehow associated with the EMT component, \(\langle T^z_{\:z}\rangle\), which in the conformal case is
related to one-third of the internal energy. By contrast, the optical description
gives the physical Polyakov-loop effective potential after the optical
volume element is converted to the Rindler one. Using this physical
Polyakov-loop effective potential, we find that real acceleration favors the
center-broken perturbative minimum and hence strengthens the tendency toward deconfinement.
We also show that the Polyakov-loop effective potential exhibits an acceleration-induced cusp and this nonanalyticity at the
perturbative minimum makes the usual curvature definition of the Debye
screening mass ill-defined. Finally, we study the analytic continuation
to imaginary acceleration and find a perturbatively confined phase in close
analogy with the imaginary-rotation case~\cite{Chen:2022smf}, as discussed in Sec.~\ref{sec:decon-con}.

Throughout this paper, we adopt the mostly-plus convention for the
Minkowski metric.

\section{Thermodynamics with Acceleration}\label{sec:gene-two-met}

We explain the formulation of describing matter with finite acceleration using nontrivial metric. We then discuss thermodynamics, particularly, the computation of the partition function from which all thermodynamic quantities are derived.  This is an exercise before introducing the Polyakov-loop background and studying the acceleration effect on the deconfinement phase transition. We find an unexpectedly subtle realization of thermodynamic properties with acceleration. We present two different approaches to accelerated systems, namely, a standard formulation based on the Rindler coordinate metric and another one based on the optical metric. Although these two formulations are supposed to be equally valid, we show that they lead to inequivalent results.

\subsection{Euclidean Rindler metric vs.\ optical metric} \label{sec:euc-opt}

First, we consider the Rindler coordinate system.  This treatment gives a plain physical interpretation that the matter and the coordinate system are co-accelerated, so that the matter looks static for the observer in this system. 

\begin{figure}
    \includegraphics[width=0.65\columnwidth]{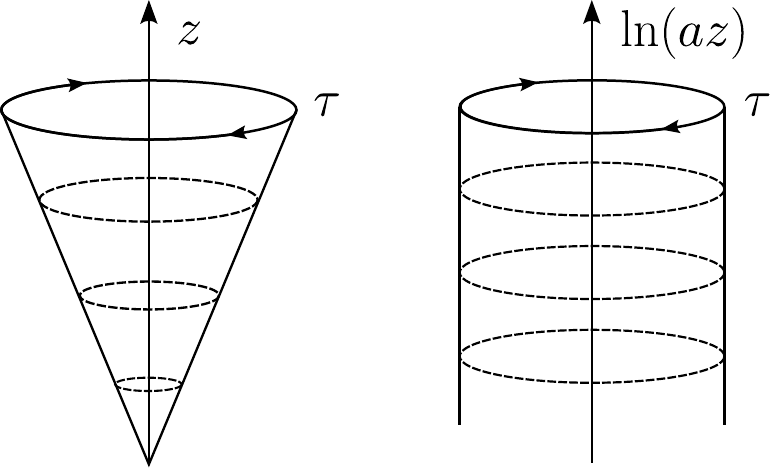}
    \caption{Schematic geometrical structures corresponding to the Euclidean Rindler metric (left) and the optical metric (right).}
    \label{fig:geometry}
\end{figure}

For the Minkowski coordinates, $(x_{\text{M}},y_{\text{M}},z_{\text{M}},t_{\text{M}})$, we consider the acceleration along the $z_{\text{M}}$-axis, and then it is convenient to introduce the Rindler frame, that is,
\begin{equation}
    t_{\text{M}} = z\sinh(a t) \,,
    \qquad 
    z_{\text{M}} = z\cosh(a t) \,, 
\end{equation}
and $x_{\text{M}}=x$ and $y_{\text{M}}=y$ where $a$ represents the proper acceleration.  We will focus on $z>0$ (right Rindler wedge) and we note that $z=a^{-1}$ should hold along the observer's worldline, which is required to take the $a\to 0$ limit properly.  Then, we can immediately derive the line element given by 
\begin{equation}
    ds_{\text{Rindler}}^2 = -(a z)^2 dt^2 + dx^2 + dy^2 + dz^2 \,. 
\end{equation}
Since we are interested in thermodynamics in the imaginary-time Matsubara formalism, we should perform all calculations with Euclidean time $\tau$, and thus, the Euclidean Rindler metric, $g_{\text{ER}}$, reads: 
\begin{equation}
    ds_{\text{ER}}^2 = (g_{\text{ER}})_{\mu\nu} dx^\mu dx^\nu 
    = (az)^2 d\tau^2 + dx^2 + dy^2 + dz^2 \,, 
\end{equation}
with $\tau$ compactified to $(0,\beta)$. Since $(\tau,z)$ has the same metric structure as the two-dimensional polar coordinates, $ds^2=dr^2 + r^2 d\theta^2$, if $\theta=a\tau$, the thermal circle along the $\tau$ direction has no singularity as long as the thermal period is $\beta=2\pi/a$.  In other words, a conical singularity appears for general temperatures that do not satisfy this condition;  see the left illustration in Fig.~\ref{fig:geometry}.

Next, we move to another spacetime via conformal transformation:
$g_{\mu\nu} \to \tilde{g}_{\mu\nu} = e^{2\sigma(x)} g_{\mu\nu}$.  Such conformal transformations from the Rindler or other curved spacetimes are often useful~\cite{Gusev:1998rp,Gibbons:1976pt,Dowker:1978md,Emparan:1994qa,Bytsenko:1995ym,Cognola:1997dv,Sonego:2010vy}.  We specifically consider the so-called optical metric that is \textit{ultrastatic} with the following properties: $g_{00} = 1$, $g_{0i} = 0$, and $g_{ij}$'s are time independent.  Therefore, the optical spacetime with acceleration $a$ is obtained by $\sigma(x) = -\ln(az)$, leading to 
\begin{equation}
  \begin{split}
    ds^2_{\text{opt}} &= (g_{\text{opt}})_{\mu\nu} dx^\mu dx^\nu
  = d\tau^2 + h_{ij} dx^i dx^j \,,\\
    h_{ij}(x) &= \frac{1}{(az)^2}\,\delta_{ij} \,.
  \end{split}
  \label{eq:g_opt}
\end{equation}
In this case, the Rindler horizon at $z=0$ is shifted to $\ln(az)|_{z=0}\to -\infty$, which should be understood from $dz^2/z^2 = [d\ln(az)]^2$; see the right illustration in Fig.~\ref{fig:geometry}.  The whole spatial part is the upper-half-space representation of hyperbolic space $H_{1/a}^3$ where the subscript indicates the curvature radius $1/a$.  Hence, the optical geometry is
\begin{equation}
    S^1_\beta \times H^{3}_{1/a} \,.
\end{equation}
In this metric, the light paths coincide with the geodesics of the spatial metric; hence the geometry is called optical.

With this optical metric, all Christoffel symbols with the $\tau$ index vanish.  This implies that, in ultrastatic spacetime, gravitational forces vanish except inertial forces of the type, $\Gamma^i_{jk} \dot{x}^j \dot{x}^k$, which leads to the existence of a covariantly constant timelike vector, $\eta^\mu$, satisfying $\nabla_\mu \eta^\nu = 0$.  This vector defines a time flow free from acceleration, rotation, and deformation, so that one can introduce a globally uniform temperature $\beta_0$ by setting $\beta^\mu = \beta_0\eta^\mu$.

In general, conformal transformations are accompanied by extra terms in the effective action, which originate from the Jacobian of the transformation. Nevertheless, on a static manifold, this Jacobian does not affect the thermal quantities derived from the partition function.  The partition function, and hence thermodynamic quantities derived from it, remain unchanged up to irrelevant terms independent of the temperature and chemical potential~\cite{Gusev:1998rp,Cognola:1997dv,Dowker:1989gp}.  Strictly speaking, this statement has not been proved in the presence of singularities, such as the conical singularity appearing in finite-temperature Euclidean Rindler space (see Fig.~\ref{fig:geometry}).  Indeed, the Euclidean Rindler geometry is flat in the bulk and is singular only at the horizon \(z=0\) for a non-Unruh temperature, whereas the corresponding optical geometry is smooth in the bulk but has constant negative spatial curvature.  The Weyl factor relating the two metrics is singular at the horizon and maps the Rindler horizon to the conformal boundary of hyperbolic space.  In this way, although thermodynamic quantities of the Rindler system are quite often analyzed in the optical system, one should be careful when comparing local densities, surface terms, and thermodynamic potentials in these two descriptions.  In the following, we will see that the optical analysis nevertheless provides the desired effective potential.

\subsection{Subtle relations between two approaches}

We will explicitly compute the partition function, $\ln Z$, using the coordinate systems mentioned above. We shall find that the results from these approaches are different, i.e.,
\begin{equation}
  \ln Z_{\text{opt}} \neq \ln Z_{\text{ER}} \,.
\end{equation}
Thus, we have different results for thermodynamic quantities such as pressure, effective potential, and so on.  This fact has previously been noted as a puzzling phenomenon that occurs in the calculations made in the Rindler spacetime~\cite{Dowker:1994fi,Moretti:1996wd}. We also refer to a recent study~\cite{Ambrus:2025dca} which has considered a fermionic system with rotation and acceleration and has pointed out a nontrivial relation between pressure from the partition function and that constructed from the energy-momentum tensor (EMT).
 
To clarify which corresponds to the physical thermodynamics, we will make a comparison to the EMT carefully.  We shall specifically consider the Hilbert EMT and its expectation value, $\expval{T^{\mu\nu}(x)}_{\text{ER}}$, given by the operator,
\begin{equation}
  T^{\mu\nu}(x) = \frac{2}{\sqrt{g_{\text{ER}}}}\,\frac{\delta S}{\delta (g_{\text{ER}})_{\mu\nu}(x)} \,.
\end{equation}
In our convention, these quantities are all defined in Euclidean Rindler spacetime.  We will see that the optical partition function satisfies the following relation:
\begin{equation}
  \frac{\partial \ln Z_{\text{opt}}}{\partial\beta_0} = -\int d^3x\, \sqrt{g_{\text{ER}}}\, \expval{T^\tau_{\:\:\tau}}_{\text{ER}} \,.
  \label{eq:OP_Ttt}
\end{equation}
This is nothing but the standard thermodynamic relation; $\partial(\beta_0 F)/\partial\beta_0 = F-T_0 \partial F/\partial T_0 = \mathcal{E}$ with the free energy $F$ and the internal energy $\mathcal{E}$
\textit{if} $\ln Z_{\mathrm{opt}}=-\beta_0 F$ holds.  This standard relation is derived immediately from the definition of the partition function, i.e.,
\begin{equation}
    Z = \Tr e^{-\beta_0 Q}\,,
\end{equation}
where $Q$ is the conserved charge with the symmetric EMT given by 
\begin{equation}
    \beta_0 Q = \int d^3x \sqrt{g} \, g_{\mu\nu} T^{\tau \mu} \beta^\nu = \beta_0\int d^3x \sqrt{g} \, T^{\tau}_{\:\:\tau} \,.
\end{equation}
It is clear that this charge, $Q$, in the accelerated coordinate system is a counterpart of the Hamiltonian for static matter.  Indeed this $Q$ can be derived in an alternative picture in which the observer sees accelerated matter in the flat coordinate system.  A short calculation shows that $Q$ corresponds to the shifted Hamiltonian, $\hat{H}_0 - a\hat{K}$, where $\hat{H}_0$ is the Hamiltonian and $\hat{K}$ is the boost operator. 
We note that $\beta(x)$ is the local inverse temperature at $x$, and $\beta_0$ is the inverse temperature of matter at rest.  More specifically, the thermodynamics in curved spacetime is well characterized by the four-temperature vector, $\beta^\mu = \beta(x)\, u^\mu$ with the fluid velocity $u^\mu$~\cite{Becattini:2016stj}.  For static matter, $\beta^\mu = \beta_0(0,0,0,1)^T$, and for accelerated matter, the local inverse temperature is given by
\begin{equation}
  \beta(x) = (a z)\,\beta_0\,, \qquad
  \beta_\mu \beta^\mu = \qty[\beta(x)]^2\,.
\end{equation}
The local inverse temperature at the reference point $z = a^{-1}$ corresponds to the global inverse temperature, $\beta_0$.  In this work, we sometimes use an alternative notation:
\begin{equation}
    T_{\text{loc}} = \frac{T_0}{az} \,,
\end{equation}
to mean the local temperature.    

Because the derivation is so simple, one may think that the partition function directly calculated in the Euclidean Rindler metric should also satisfy Eq.~\eqref{eq:OP_Ttt}.  However, surprisingly, $\ln Z_{\text{ER}}$ is inconsistent with this expected relation, but instead, we find that it is related to the EMT through
\begin{equation}
  \ln Z_{\text{ER}} = -\beta_0\int d^3x\, \sqrt{g_{\text{ER}}} \expval{T^z_{\:\:z}}_{\text{ER}}\,.
  \label{eq:ER_Tzz}
\end{equation}
We do not yet have full understanding of why the partition function in the Euclidean Rindler coordinates deviates from the physical partition function.  Nevertheless, the above relation has been confirmed in an independent calculation~\cite{Ambrus:2025dca}.  These subtle relations hold even in the case with the Polyakov-loop background.

In this work, we aim to figure out a physically correct effective potential, $V_{\text{eff}}(x)$, and compute the Polyakov-loop expectation value by minimizing it.
It is natural to define the local effective potential from the integrand of the partition function:
\begin{equation}
    \int d^4x\, \sqrt{g_{\text{ER/opt}}}\, V_{\text{ER/opt}}(x) = -\ln Z_{\text{ER/opt}}\,.
\end{equation}
Recalling that these \textit{different} partition functions are related to the EMT components, we can find a relation between them.  Specifically, the energy-momentum conservation law gives
\footnote{In this paper, we always work in Euclidean Rindler spacetime when we refer to the EMT\@.  Therefore, $T^{zz}$ and $T^z_{\:\:z}$ are identical, while $T^{\tau\tau} = g_{\text{ER}}^{\tau\tau} T^\tau_{\:\:\tau} = (az)^{-2} T^\tau_{\:\:\tau}$.}
$\nabla_\mu \expval{T^{\mu z}}_{\text{ER}} = (\partial_z + \Gamma^\tau_{\tau z}) \expval{T^{zz}}_{\text{ER}} + \Gamma^z_{\tau\tau} \expval{T^{\tau\tau}}_{\text{ER}} = 0$, that is translated into $\partial_z (z \expval{T^{zz}}_{\text{ER}}) - (az)^2 \expval{T^{\tau\tau}}_{\text{ER}} = 0$. According to the mass dimension of $\expval{T^{zz}}_{\text{ER}}$, it is found to be proportional to $z^{-4}$ as we will see in Eq.~\eqref{eq:Tzz_ER}.  Thus, we can conclude that $(az)^2 \expval{T^{\tau\tau}}_{\text{ER}}=\expval{T^\tau_{\:\:\tau}}_{\text{ER}} = -3\expval{T^z_{\:\:z}}_{\text{ER}}$.  Using these relations, we reach the following relation:
\begin{equation}
    \ln Z_{\text{opt}} = -3\int^{\beta_0}\frac{d\beta}{\beta}\, \ln Z_{\text{ER}}\,,
\end{equation}
which can be translated into the relation between the integrands as
\begin{equation}
  \sqrt{g_{\text{opt}}}\, V_{\text{opt}} = - \frac{3}{\beta_0} \int^{\beta_0} d\beta \sqrt{g_{\text{ER}}}\, V_{\text{ER}} \,.
\end{equation}
Recalling $(g_{\text{opt}})_{\mu\nu}=e^{2\sigma(x)} (g_{\text{ER}})_{\mu\nu}$ and thus $\sqrt{g_{\text{opt}}}=(az)^{-4} \sqrt{g_{\text{ER}}}$, it would be natural to define the physical local effective potential as
\begin{equation}\label{eq:Vrelation}
    V_{\text{eff}}(x) = (az)^{-4}\, V_{\text{opt}} = -\frac{3}{\beta_0} \int^{\beta_0} d\beta\, V_{\text{ER}}(x)\,.
\end{equation}
As we see later, $V_{\text{opt}}$ has no spatial dependence, and the $z$ dependence in $V_{\text{eff}}(x)$ is factorized into $e^{4\sigma(x)}=(az)^{-4}$ in the above definition.

\section{Polyakov-Loop Effective Potential}\label{sec:poly-pot}

We shall consider confinement-deconfinement properties in accelerated systems by performing the one-loop calculation of the Polyakov-loop effective potential at high temperature, which is often called the Gross-Pisarski-Yaffe-Weiss potential~\cite{Gross:1980br,Weiss:1980rj,Weiss:1981ev}. In our calculations, the gauge field is decomposed into the background part and the dynamical part. The former corresponds to the gluonic contributions at high temperature limit, and it has only a diagonalized $A_{\text{B}\tau}$ background part 
\begin{equation}
    A_{\text{B}\tau} =\frac{1}{g\beta}\,\bm{\phi}\cdot\bm{H} \,, 
\end{equation}
where $\bm{H}$ is a vector of basis elements of the Cartan subalgebra $\mathfrak{h}$ of $\su(N_c)$ in color space.  Here, parameter $\bm{\phi}$ describes $A_{\text{B}\tau}$ degrees of freedom. For $SU(N_c)$, $\bphi$ is a vector with $N_c-1$ components. Since the leading order of the Polyakov loop is represented by this parameter as 
\begin{equation}\label{eq:L_def}
    L = \inv{N_c} \Tr \exp\qty[ig\int_0^\beta A_{\mathrm{B}\tau}\,ds] = \inv{N_c} \Tr e^{i\bm{\phi}\cdot\bm{H}} \,,
\end{equation}
we shall often call the parameter $\bphi$ the Polyakov-loop background parameter. The effective potential of the system is calculated at the one-loop order of the dynamical gauge part, and the resulting effective potential is a function of the Polyakov loop; we shall refer to this as the Polyakov-loop effective potential. The minimum locations of the Polyakov-loop effective potential give us information about confinement-deconfinement properties in accelerated systems.

We also note our definition of the covariant derivative. In a curved spacetime, we have a covariant derivative of both gauge field and spacetime. We shall use $G_\mu=\nabla_\mu+igA_\mu$ to describe the combination of these two derivatives, where the temporal component has a background gauge field as
\begin{equation}
    D_{\mathrm{B}\tau} = \partial_\tau + i\frac{\bphi\cdot\bm{H}}{\beta}\,,
    \qquad
    G_{\mathrm{B}\tau} = \nabla_\tau + i\frac{\bm{\phi}\cdot\bm{H}}{\beta}\, .
    \label{eq:cov_derivatives}
\end{equation}
Here, for later convenience, we introduced the covariant derivative, $D_{\mathrm{B}\tau}$, without connection in the spacetime part.

\subsection{Calculations with the Rindler Coordinate Metric}

In this work, we focus on the right Rindler wedge, so that $z > 0$ is always chosen.  The one-loop gluonic action in a curved spacetime can be written in terms of the covariant derivative, $G_\mu$, and the eigenfunctions and eigenspectra provide the thermodynamics.  We set the gauge fixing condition to $G_\mu A^\mu = 0$, and the one-loop effective potential takes the form of the determinant of the covariant derivatives under acceleration, that is, the determinant of the scalar and vector Laplacian operators.  The scalar ghost contribution is given by the determinant of
$- \GBs^2 = - (az)^{-2}D_\tau^2 - \partial_x^2 - \partial_y^2 - \partial_z^2 - z^{-1}\partial_z$.
This contribution exactly cancels non-physical modes in the gluonic sector.  For the gluon modes, we simplify the notation of the vector Laplacian operator by introducing a $2\times2$ submatrix, $L_{\text{v}}$ as
\begin{equation}
    \qty(\Lv)^\mu_{\:\: \nu} =
    \begin{pmatrix}
    - \GBs^2 + z^{-2} & 2z^{-1}\,D_\tau\\[3pt]
    -2a^{-2} z^{-3}\, D_\tau & -\GBs^2 - 2z^{-1}\partial_z
    \end{pmatrix} \,,
    \label{eq:Lv}
\end{equation}
which gives
\begin{equation}
    - G_{\mathrm{B\,v}}^2 =
    \begin{pmatrix}
      - \GBs^2 & 0 & 0 \\
      0 & - \GBs^2 & 0 \\
      0 & 0 & L_{\mathrm{v}}
    \end{pmatrix} \,.
    \label{eq:LaplacianV}
\end{equation}
Then, we can express the partition function in terms of only the submatrix contribution:
\begin{equation}
    \ln Z_{\text{ER}} = \ln\Det\qty(-\GBs^2) - \inv{2} \ln\Det\qty(-G_{\mathrm{B, v}}^2) = - \inv{2} \Tr \ln \qty(\Lv) \,.
\end{equation}
We can perform this calculation in a direct method but here let us first present a slightly tricky derivation in which a possible connection to $\expval{T^{zz}}$ becomes transparent.  We will perform the more direct calculation later to confirm that two calculation schemes give the same answer.

\subsubsection{Derivation from the EMT}

We shall introduce a transient variation of the metric by hand; $g_{zz} = e^{2\sigma(z)}$, which would not change the results if we set $\sigma \to 0$ at the end.  Instead of computing $\ln Z$ directly, it would be more instructive to compute the variation, $\delta \ln Z$, in response to $\delta\sigma(z)$.  Because $\delta\sigma(z)$ depends on $z$, three contributions arise as
\begin{align}
    - \delta \ln Z 
    &= \int d^4x \, \sqrt{g}\, g_{zz}\expval{T^{zz}} \, \delta\sigma
      = \frac{1}{2} \Tr \qty[\Lv^{-1} \delta \Lv] \,, \\
    \delta \Lv &= \frac {\partial \Lv}{ \partial \sigma } \delta\sigma + \frac {\partial \Lv}{ \partial \sigma' } \delta\sigma' + \frac {\partial \Lv}{ \partial \sigma'' } \delta\sigma'' \,. 
\end{align}
Here, $\sigma'=\partial\sigma/\partial z$ and $\sigma''=\partial^2\sigma/\partial z^2$.  We convert $\delta\sigma'$ and $\delta\sigma''$ into $\delta\sigma$ by the integration by parts and then obtain the expression involving $\expval{T^{zz}}$.  The trace calculation needs the physical modes, i.e., the eigenfunctions of the submatrix $L_\mathrm{v}$. These modes are represented in terms of a scalar eigenfunction $\Phi$ as
\begin{subequations}
  \begin{align}
    A_{1\,i=\{z,\tau\}} &= \frac{1}{\lambda}\qty(\partial_z\Phi,\, D_\tau\Phi)\,, \\
    A_{2\,i=\{z,\tau\}} &= \frac{1}{\lambda}\qty(-\inv{az}D_\tau\Phi,\, az\partial_z\Phi)\,.
  \end{align}
  \label{eq:A_physical}
\end{subequations}
Here, $\lambda$ is a label for the complete basis functions.  If we need $A_{1,2}^i$, we should carefully multiply $(g_{\text{ER}})^{\tau\tau}=(az)^{-2}$ to the $\tau$ component.  The scalar eigenfunction is given as 
\begin{equation}
  \Phi_{n,\lambda, \bkt}(x) = \frac{E_{\balpha}}{2\pi}\sqrt{\frac{\lambda}{a\beta_0}} e^{i\omega_n\tau+i\bkt\cdot\bxt}\, J_{\abs{\nu_n}}(\lambda z) \,,
  \label{eq:Phi_mode}
\end{equation}
where the background shifted Matsubara frequency is defined as
\begin{equation}
  \nu_n = a^{-1} \qty( \omega_n + \frac{\bm{\phi}\cdot \balpha}{\beta_0} ) = n v + \bar{\phi} \,.
  \label{eq:shifted_Matsubara}
\end{equation}
where \(\balpha\) denotes an element of the root system or a zero root of
\(\mathfrak{su}(N_c)\), and $E_{\balpha}$ is the corresponding generator. Here, we introduced $\tia = a \beta_0$ and $v=2\pi/\tia$.
The first term corresponds to the Matsubara frequency, $\omega_n=2\pi n \beta_0^{-1}$ made dimensionless with $a$.  From this expression, it is evident that $\bm{\phi}\cdot\balpha>2\pi$ can be absorbed in the increment of $n$.  Therefore, without loss of generality, we can limit the Polyakov-loop background as $\bar{\phi}=\tia^{-1}(\bm{\phi}\cdot\balpha)_{\text{mod $2\pi$}}$.  Using the differential equation of the Bessel functions, it is easy to verify
\begin{equation}
  -\GBs^2 \Phi_{n,\lambda,\bkt} = (\lambda^2 + \bkt^2) \Phi_{n,\lambda,\bkt}
\end{equation}
as well as the normalization condition, $\int dx\, |\Phi_{n,\lambda,\bkt}(x)|^2=1$.
Then, we can confirm $L_{\mathrm{v}}\,A_{1,2} = (\lambda^2+\bkt^2)\,A_{1,2}$ as expected.  In the following, we simplify the notation as $\bkt^2 \to k^2$ as long as it is not confused with the four-vector squared.

Now, we are ready to proceed to the trace calculations.  It is convenient to replace $(\lambda^2+k^2)^{-1}$ in the trace with the exponential integral form, that is,
\begin{equation}
  \begin{split}
  &\inv{2} \Tr\qty[\Lv^{-1} \, \delta \Lv] \\
  & =\int d^4x \, \sqrt{g_{\text{ER}}} \, \sum_{\balpha, n} \int \frac{kdk}{2\pi \tia}\, \int_0^\infty \frac{d\lambda}{\lambda} \int_0^\infty \! ds\, e^{-s(\lambda^2+k^2)} \delta\Lambda_\sigma \,.
  \end{split}
\end{equation}
It is easy to see that the $s$ integration recovers the eigenvalue of $\Lv^{-1}$.  We also note that $\lambda$ and $\tia$ in the denominator appear from the normalization of the basis functions.  The integrand $\delta\Lambda_\sigma$ corresponds to $\delta\Lv$ and its explicit form is
\begin{equation}
  \begin{split}
    \delta\Lambda_\sigma &= 2\qty(\frac{\nu^2}{z^3}JJ' \!+\! \frac{\nu^2}{z^2} JJ'' \!-\! \frac{1}{z^2} J'J' \!+\! \frac{1}{z} J'J'' \!+\! J'J''')\delta\sigma \notag\\
    &\quad - \inv{2} \qty( J'J' + \frac{\nu^2}{z^2}J^2 )\qty(\inv{z} \delta\sigma' + \delta\sigma'') \,,
  \end{split}
\end{equation}
where we simplify the notations as $\nu_n \to \nu$, $J_{|\nu|}(\lambda z) \to J$, and $\partial_z J \to J'$, etc.  Here, the important step for further proceeding to the calculation is that we take the $\lambda$ integration first and then the $s$ integration last.  As a result of the $\lambda$ integration, we find the kernel functions defined by
\begin{align}
  K_\nu (s; z,z') &= \int_0^\infty \,kdk\,\int_0^\infty  \lambda d\lambda\, e^{-s\qty(\lambda^2+k^2)} J_\nu(\lambda z) J_\nu(\lambda z') \no\\
  &= \inv{4s^2} e^{-\frac{z^2+z^{\prime2}}{4s}} I_\nu\qty(\frac{zz'}{2s}) \,.
  \label{eq:kernel}
\end{align}
Here, $\nu$ can be any real number, and $I_\nu(x)$ is the modified Bessel function of the first kind.  For this trace calculation, there appear many terms, but after the integration by parts to extract $\delta\sigma(x)$ in the integrand, only one term is left as
\begin{equation}
  \expval{T^{zz}}_{\text{ER}}
  = -\sum_{\balpha, n} \inv{2\pi \tia} \int \frac{ds}{s}  K_{|\nu_n|}(s;z,z) \,.
  \label{eq:Tzz_integral}
\end{equation}
Now, the remaining tasks are taking the Matsubara frequency sum and then the $s$ integration.  For technical steps, see Appendix~\ref{app:der_Tzz}.  After all, we arrive at the following expression in the presence of the Polyakov-loop background:
\begin{equation}
  \begin{split}
  &\expval{T^{zz}}_{\text{ER}} \\
  &= \sum_{\balpha} \frac{1}{z^4}\qty[ \frac{2\pi^2}{3\tia^4}\mathrm{B}_4\qty(\frac{\bm{\phi}\cdot \bm{\alpha}}{2\pi}) - \frac{1}{3\tia^2}\mathrm{B}_2\qty(\frac{\bm{\phi}\cdot \bm{\alpha}}{2\pi}) + \frac{11}{720\,\pi^{2}} ] \,,
  \end{split}
  \label{eq:Tzz_ER}
\end{equation}
where $\bm{\phi}\cdot \bm{\alpha}\in[0,2\pi)$ (mod $2\pi$ is taken implicitly), and $\mathrm{B}_n(x)$ denotes the Bernoulli polynomial of order $n$.  Specifically,
\begin{equation}
  \begin{split}
  \mathrm{B}_2(x) &= x^2 - x + \frac{1}{6}\,,\\
  \mathrm{B}_4(x) &= x^4 - 2x^3 + x^2 - \frac{1}{30}\,.
  \end{split}
\end{equation}
In the special limit of $\bphi=0$, only the constant terms in $\mathrm{B}_n(x)$ remain nonvanishing, and we recover the one-loop EMT expression for hot gluonic matter, that is,
\begin{equation}
  \expval{T^{zz}}_{\text{ER}} \biggr|_{\bar{\phi}=0}  = \frac{\Nc^2-1}{z^4}
    \left(
    -\frac{\pi^2}{45 \tia^4}
    -\frac{1}{18 \tia^2}
    +\frac{11}{720\pi^2}
    \right) \,.
    \label{eq:VgWOSurface}
\end{equation}
This result agrees with previous calculations of the EMT for electromagnetic fields~\cite{Lapygin:2025zhn,Dowker:1994fi,Frolov:1987dz,Page:1982fm}.

If we use the relation in Eq.~\eqref{eq:ER_Tzz} (that is not yet derived but will be discussed below), we get the following form of the Polyakov-loop effective potential:
\begin{align}
  V_{\text{ER}}
  &= T_{\text{loc}}^4 \sum_\balpha  \qty[ \frac{2\pi^2}{3} \mathrm{B}_4\qty(\!\frac{\bphi\cdot\balpha}{2\pi}\!) \!-\! \frac{\tilde{a}^2}{3} \mathrm{B}_2\qty(\!\frac{\bphi\cdot\balpha}{2\pi}\!) \!+\! \frac{11 \tilde{a}^4}{720\pi^2} ] \no \\
  &= T_{\text{loc}}^4 \sum_\balpha \biggl[ -\frac{\pi^2}{45} - \frac{\tilde{a}^2}{18} + \frac{(\bphi\cdot\balpha)^4}{24\pi^2} - \frac{\abs{\bphi\cdot\balpha}^3}{6\pi} \no \\
  &\quad + \qty(\inv{6} - \frac{\tilde{a}^2}{12\pi^2})(\bphi\cdot\balpha)^2 + \frac{\tilde{a}^2}{6\pi}\abs{\bphi\cdot\balpha} +\frac{11 \tilde{a}^4}{720\pi^2} \biggr] \,.
       \label{eq:VPol_ER}
\end{align}
We will see that this is obtained from the direct calculation.  We can check that the zero-acceleration limit is reasonable.  Evaluating the potential on the observer's worldline by setting $z=1/a$, we can take the $a\to 0$ limit, in which only the $\mathrm{B}_4$ term survives.  In this way, we recover the well-known Gross-Pisarski-Yaffe (GPY) or Weiss potential~\cite{Weiss:1980rj,Weiss:1981ev,Gross:1980br,Fukushima:2017csk} correctly.  However, Eq.~\eqref{eq:VPol_ER} is \textit{not} the physically correct answer.  As we discuss later, the difference from the correct answer lies in the $\mathrm{B}_2$ term, and thus, the zero-acceleration limit does not impose any constraint. 

\subsubsection{Derivation from the direct calculation}

With the solution of physical modes in Eq.~\eqref{eq:A_physical}, the local thermodynamic potential can be expressed in terms of the corresponding heat-kernel representation,
\begin{equation}
    \begin{split}
        V_{\text{ER}}
        &= -\sum_{\balpha,n} \int d^2\bm{k}\int_0^\infty d\lambda
        \int_0^\infty \frac{d s}{s}\,
        e^{-s(\lambda^2 + k^2)} \no\\
        &\qquad \times
        \frac{\partial_z \Phi\partial_z \Phi^* + (az)^{-2}(D_\tau \Phi)(D_\tau \Phi)^*}{\lambda^2} \,.
    \end{split}
\end{equation}
We can perform an integration by parts and use the equation of motion for \(\Phi\), which yields:
\begin{equation}
    \begin{split}
        V_{\text{ER}}
        =& -2\pi\sum_{\balpha,n} \int_0^\infty k dk\,\int_0^\infty d\lambda\,
        \int_0^\infty \frac{d s}{s}\,
        e^{-s(\lambda^2 + k^2)} \\
        &\qquad\qquad\quad \times
        \biggl[ |\Phi|^2 + \frac{1}{\lambda^2 z}\partial_z(z\Phi^*\partial_z\Phi) \biggr]\,.
    \end{split}
\end{equation}
The second term in the brackets becomes a total derivative when integrated
with the Euclidean Rindler measure \(\int d^4x\,\sqrt{g_{\rm ER}}\).  It
therefore contributes only through a surface term.  As shown in
Appendix~\ref{app:surface}, this surface contribution can be related to the
trace of the minimal scalar energy-momentum tensor. Such a surface contribution is gauge dependent and may lead to a negative contribution to the entropy, as discussed in Refs.~\cite{Iellici:1996jx,Kabat:1995eq}.
In the present work, we discard this surface term and retain only the bulk contribution, leading to
\begin{equation}
  \begin{split}
    V_{\text{ER, bulk}}
    &= -\sum_{\balpha, n} \int_0^\infty \frac{k dk}{2\pi\tia} \int_0^\infty\lambda d\lambda \\
    &\qquad\qquad \times \int_0^\infty \frac{d s}{s}\,
    e^{-s(\lambda^2 + k^2)} \, J_{\nu_n}^2(\lambda z) \,.
  \end{split}
  \label{eq:VgWOsurface_integral}
\end{equation}
In view of Eqs.~\eqref{eq:kernel} and \eqref{eq:Tzz_integral} with $z'=z$, this bulk contribution is just equivalent to $\expval{T^{zz}}_{\text{ER}}$.

\subsection{Calculations with the Optical Metric}

Let us move to the optical metric calculation.
Considering the mathematical analogy between density and the background gauge field $\bphi$, and the fact that $\bphi$ appears only at finite temperature since it can be absorbed by the gauge transformation at zero temperature; it seems that the difference in the Polyakov-loop effective potential before and after the conformal transformation (the Jacobian of this conformal transformation) does not contain $\bphi$ dependence. However, a proof of this statement is beyond the scope of the present
work.
Nevertheless, the results obtained in the optical spacetime undoubtedly provide significant insights into the effective potential in the Rindler spacetime.

Before discussing the optical metric calculation, it is useful to see the goal in advance.  Because we already know $\expval{T^{zz}}$, we can read $\expval{T^\tau_{\:\:\tau}} = -3\expval{T^z_{\:\:z}}$, with which we can perform the $\beta_0$ integration to find the effective potential using Eq.~\eqref{eq:OP_Ttt}.  The term $\propto \mathrm{B}_4$ is not changed since it is proportional to $\beta_0^{-4}$, and its $\beta_0$ integration leads to a factor of $-1/3$ which is canceled by $-3$ of the EMT coefficient. The difference from the heat-kernel calculation in Rindler spacetime appears in the term proportional to $\mathrm{B}_2$. The expected physical form of the effective potential reads:
\begin{align}
  V_{\text{opt}}
  &= T_0^4 \sum_\balpha \qty[ \frac{2\pi^2}{3} \mathrm{B}_4\qty(\!\frac{\bphi\cdot\balpha}{2\pi}\!) \!-\! \tilde{a}^2 \mathrm{B}_2\qty(\!\frac{\bphi\cdot\balpha}{2\pi}\!) \!-\! \frac{11 \tilde{a}^4}{240\pi^2} ] \no \\
  &= T_0^4 \sum_\balpha \biggl[ -\frac{\pi^2}{45} - \frac{\tilde{a}^2}{6} + \frac{(\bphi\cdot\balpha)^4}{24\pi^2} - \frac{\abs{\bphi\cdot\balpha}^3}{6\pi} \no\\
  &\quad + \qty(\inv{6} - \frac{\tilde{a}^2}{4\pi^2})(\bphi\cdot\balpha)^2 + \frac{\tilde{a}^2}{2\pi}\abs{\bphi\cdot\balpha} -\frac{11 \tilde{a}^4}{240\pi^2} \biggr] \,.
        \label{eq:VPol_OP}
\end{align} 
As before, $\bphi\cdot\balpha\in [0,2\pi)$ (mod $2\pi$ is taken implicitly). The \(\tilde a^4\) term is fixed by the thermodynamic relation Eq.~\eqref{eq:Vrelation} rather than by
the optical heat-kernel expansion used below.  Since the \(\tilde a^4\) term is independent of the Polyakov-loop background \(\phi\), the choice of this term does not affect the minimization of the effective potential. 

\subsubsection{Formal expressions}

In the optical spacetime, we will utilize the heat kernel method.  The spatial slice $(\Sigma^3,h_{ij})$ is the hyperbolic space $H^3_{1/a}$ of
radius $L=1/a$ in the Poincar\'e half-space representation, with which the curvature is $R = -6a^2$, where $a$ is the acceleration of the corresponding Rindler spacetime.
As we already emphasized, as a characteristic of the ultrastatic metric, all Christoffel symbols involving the $\tau$ index vanish, so the overall curvature of the optical spacetime is equal to the curvature of three-dimensional hyperbolic space. 
This fact is convenient when employing the heat kernel method: the Laplace-type operator $D_4 = -(\Delta_4 + E)$ in four-dimensional spacetime (where $E$ denotes some general matrix depending on specific problems) can be split into the time and spatial components as
\begin{equation}
    D_4 = -\qty[g^{\mu\nu}\nabla_\mu \nabla_\nu + E(x)]
    = - \partial_\tau^2 + D_3 \,,
\end{equation}
where $\nabla_\mu$ represents the covariant derivative in curved spacetime, and since $\partial_\tau$ and $D_3$ commute with each other, the heat kernel at finite temperature can also be decomposed as
\begin{equation}
    \Tr \ln D_4 = - \int_0^\infty \frac{ds}{s} \Tr e^{-s D_4}
     = - \int_0^\infty \frac{ds}{s} K_\tau (s) \, K_\Sigma (s) \,,
\end{equation}
where
\begin{align}
  K_\tau (s) &= \sum_{n} e^{-s \omega_n^2} \,,
  \label{eq:Ktau}\\
  K_\Sigma (s) &= \frac{1}{(4\pi s)^{3/2}} \sum_{m=0}^\infty s^m 
  \int d^3x\, \sqrt{\abs{h_{ij}}}\, \tr a_m(x)
  \label{eq:KSigma}
\end{align}
with the Matsubara frequency $\omega_n$ and Seeley--DeWitt coefficients $a_m(x)$ (see, e.g., Ref.~\cite{Vassilevich:2003xt} for the definitions and expressions).  Strictly speaking, we need some regularization to obtain this formula, and we shall explain it later.

For the spatial part, we shall consider the first two coefficients so that we can pick up acceleration dependence at the leading order. These terms correspond to the $T_0^4$ and $a^2T_0^2$ dependencies in the effective potential.
The coefficients $a_m(x)$ in the present setup are given by
\begin{equation}
    a_0(x) = \bm{1} \,, \qquad a_1(x) = E(x) + \inv{6} R(x)\, \bm{1}\,.
\end{equation}
The unity here is the identity matrix of the size associated with the operator; in this case, it is of size $1\times1$ for scalars, while $4\times4$ for vectors.
The next leading order yields a term proportional to $a^4$, which has either dependence on $\ln T_0$ or no $T_0$ dependence at all.  In fact, a short calculation shows that the third coefficient $a_2$ cancels out between the ghost and the gluon in our scheme, so our calculation determines the effective potential in the optical spacetime up to the $a^4$ order. Nevertheless, when considering the effective potential in the Rindler spacetime, terms independent of $T_0$ may arise from the Jacobian associated with the conformal transformation. Such terms correspond to an offset of the effective potential and are proportional to $a^4$.
Here, we calculate the effective potential in the optical spacetime to the order of $a^2 T_0^2$ and derive $V_{\text{opt}}$ by neglecting the $a^4$ dependence.

For the temporal part, in the case of the optical metric, the Christoffel symbols are trivial and we use the covariant derivative, $D_{\mathrm{B}\tau}$, as introduced in Eq.~\eqref{eq:cov_derivatives}. 
For integer-spin particles, this temporal covariant derivative shifts the Matsubara frequency as already seen in Eq.~\eqref{eq:shifted_Matsubara}.

Applying the formula
\begin{equation}
  \sum_{n} e^{-A^2(n+B)^2} 
     = \frac{\sqrt{\pi}}{A} + \frac{2\sqrt{\pi}}{A} \sum_{k=1}^\infty \cos\qty(2\pi kB)\, e^{-k^2\pi^2/A^2}
\end{equation}
for $A = 2\pi T_0\sqrt{s}$ and $B = \bphi\cdot\balpha/2\pi$,
shifting the Matsubara frequency in the presence of the Polyakov-loop background according to Eq.~\eqref{eq:shifted_Matsubara}, we find that the kernel in Eq.~\eqref{eq:Ktau} is modified as
\begin{equation}
    K_\tau(s;\bphi)
  = \frac{\beta_0}{2\sqrt{\pi s}}
    \left[
      1 + 2\sum_{k=1}^{\infty}
          \cos(k\bphi\cdot\balpha)\,
          e^{-\frac{\beta_0^2 k^2}{4s}}
    \right].
    \label{eq:temporal_heatkernel_all}
\end{equation}

For later convenience, we separate this kernel into the \textit{vacuum} and \textit{thermal} parts as
\begin{equation}
    K_{\tau,0}(s) = \frac{\beta_0}{2\sqrt{\pi s}}\,, \quad K_{\tau,\mathrm{th}}(s;\bphi) = K_{\tau}(s;\bphi) - K_{\tau,0}(s) \,.
\end{equation}
This is useful when we consider thermodynamic quantities such as the effective potential derived from $T_0 \Tr \ln D_4$. The heat kernel method requires some regularization by subtracting a part of the kernel, and when we subtract the vacuum part of the kernel for the regularization,
\begin{equation}
    T_0 \Tr \ln \qty( D_4 / D_4') = - \int_0^\infty \frac{ds}{s}\,T_0 \, K_{\tau,\mathrm{th}}(s;\bphi)\, K_\Sigma(s) \,,
\end{equation}
is satisfied for some $D_4'$. Here, from our definition, the right-hand side exhibits precisely the same dependence on $T_0$ as the case without the regularization (i.e., including all $K_\tau$). Simultaneously, this regularization also keeps $\bphi$-dependence since $K_{\tau,0}$ does not depend on $\bphi$.
Therefore, the regularization operator \(D_4'\) is totally irrelevant in our thermal calculations and we can simply drop this contribution. The right-hand side can be interpreted as the finite-temperature Polyakov-loop effective potential.
In the following, therefore, we simply write $\Tr\ln (D_4/D'_4) \to \Tr\ln D_4$.

\subsubsection{Heat kernel expansion for ghosts and gluons}
\label{sec:gluon_optical}

For the ghosts and gluons, we consider the Laplacians for them in the optical spacetime. As in the previous calculations, the gauge field is decomposed into the background part and the dynamical part as $A_\mu = A_{\mathrm{B}\mu} + \mathcal{A}_\mu$, and then the Yang--Mills action becomes
\begin{equation}
  \begin{split}
    \tr F_{\mu\nu}F^{\mu\nu} &= -2\tr \mathcal{A}_\mu \qty(G_{\mathrm{B}}^2) \mathcal{A}^\mu \\
    &\qquad + 2\tr \mathcal{A}^\nu \qty[\nabla_{\mu}, \nabla_{\nu}] \mathcal{A}^\mu + \order{\mathcal{A}^3}\,,
  \end{split}
  \label{eq:expand_D2matrix_2}
\end{equation}
where we used $F_{\mathrm{B}}^{\mu\nu} = 0$, which comes from the $A_{\mathrm{B}\tau}$ parameter condition. The second term becomes nontrivial in the optical spacetime since a finite curvature appears, i.e.,
\begin{equation}
    \mathcal{A}^\nu \qty[\nabla_{\mu}, \nabla_{\nu}]\mathcal{A}^\mu
    = \mathcal{A}_\mu R^\mu_{\;\nu} \mathcal{A}^\nu \,,
\end{equation}
where $R^\mu_{\;\nu}$ is the Ricci tensor.  Therefore, the partition function is obtained by
\begin{align}
  \ln Z_{\text{opt}} 
  = \Tr \ln \qty(-G_{\mathrm{B\, s}}^2) - \inv{2} \Tr \ln \qty[- (g_{\text{opt}})^\mu_{\;\;\nu} G_{\mathrm{B}\,v}^2 + R^\mu_{\;\;\nu}] \,.
\end{align}
We can calculate these terms using the heat-kernel formulas substituting $D_4 = -G_{\mathrm{B\, s}}^2$ and $D_4 = -g^\mu_{\;\;\nu} G_{\mathrm{B\, v}}^2 + R^\mu_{\;\;\nu}$.  Then, we have
\begin{align}
    \Tr\ln D_4
    &= -\int_0^\infty \frac{\beta_0\,ds}{8\pi^2s^3}\, 
    \sum_{k=1}^{\infty} \cos(k\bphi\cdot\balpha)\, e^{-\frac{\beta_0^2 k^2}{4s}} \no \\
    & \qquad \times \int d^3x\,\sqrt{\abs{h_{ij}}} \tr \qty[a_0(x) + sa_1(x)]
\end{align}
for ghosts and gluons.  We note that $\sqrt{|h_{ij}|}$ is nothing but $\sqrt{g_{\text{opt}}}$, which is clear from the definition in Eq.~\eqref{eq:g_opt}.
The difference between the ghost and gluon sectors lies only in the value of $a_0(x)$ and $a_1(x)$.  The first two local Seeley--DeWitt coefficients for ghost with $E = 0$ are
\begin{equation}
    a_0(x) = 1, \qquad
    a_1(x) = \frac{1}{6}R = - a^2 \,,
\end{equation}
while the coefficients for gluons with $E = -R^\mu_{\;\;\nu}$ are
\begin{equation}
    \tr a_0(x) = 4, \quad
    \tr a_1(x) = \tr \qty( -R^\mu_{\ \nu} + \frac{1}{6}R) = - \frac{R}{3} = 2a^2 .
\end{equation}

Recalling the relation, $\int d^4x \sqrt{g_{\text{opt}}}\, V_{\text{opt}}=-\ln Z_{\text{opt}}$, we can identify the respective parts of the effective potential as
\begin{align}
    V_{\mathrm{ghost}}
    &= \frac{2T_0^4}{\pi^2} \sum_\balpha \sum_{l=1}^{\infty} \biggl[ \frac{\cos(l\bphi\cdot\balpha)}{l^4} - \frac{\tia^2}{4} \frac{\cos(l\bphi\cdot\balpha)}{l^2} \biggr]
\end{align}
from the ghost contribution and
\begin{align}
    V_{\mathrm{gluon}}
    &= -\frac{4T_0^4}{\pi^2} \sum_\balpha \sum_{l=1}^{\infty} \biggl[ \frac{\cos(l\bphi\cdot\balpha)}{l^4} + \frac{\tia^2}{8} \frac{\cos(l\bphi\cdot\balpha)}{l^2} \biggr]
\end{align}
from the gluon contribution, respectively.
To derive the above results, we performed the $s$-integration using the formula:
\begin{equation}
  \int_0^\infty ds\, s^{-n-1} e^{-\frac{\beta_0^2 k^2}{4s}}
  = \left(\frac{\beta_0^2 k^2}{4}\right)^{-n} \Gamma(n) \,.
  \label{eq:heatkernel_Gamma_formula}
\end{equation}

In the first term of the potential, the ghost contribution cancels half of the gluon contribution, resulting in the well-known cancellation of non-physical modes. In contrast, in the second term (which provides the leading effect of acceleration), both contributions have the same coefficient. The ghost does not cancel the gluon effect but rather enhances the overall effect. This amplification of gluons (including non-physical modes) by ghosts is a peculiar behavior in the accelerated system.  In total, the effective potential is
\begin{align}
  V_{\text{opt}} &= V_{\text{gluon}} + V_{\text{ghost}} \no\\
  &= T_0^4 \sum_\balpha  \qty[ \frac{2\pi^2}{3} \mathrm{B}_4\qty(\frac{\bphi\cdot\balpha}{2\pi}) - \tilde{a}^2 \mathrm{B}_2\qty(\frac{\bphi\cdot\balpha}{2\pi}) ] \,,
\end{align}
which coincides with the expected form in Eq.~\eqref{eq:VPol_OP} apart from the $a^4$-order terms as mentioned.

\section{Discussions on the Polyakov-Loop Behavior} \label{sec:decon-con}

It is worth noting that the physical effective potential inferred
from the Rindler EMT through the thermodynamic relation
agrees with the Rindler-frame effective potential obtained from
the optical calculation. More explicitly, the optical calculation naturally gives the effective potential \(V_{\text{opt}}\) with respect to the optical volume element.  Rewriting the same partition function in terms of the original Rindler volume element gives
\begin{equation}
    V_{\text{eff}}(x)
    =
    e^{4\sigma(x)} V_{\text{opt}}(x) \,.
\end{equation}
This quantity, rather than \(V_{\text{opt}}\) itself, coincides with the local
thermodynamic potential obtained from the Rindler EMT calculation.

We analyze this physical effective potential.  Let us write it down here.
\begin{equation}
  \begin{split}
    &V_{\text{eff}}(x) \\
  &= T_{\text{loc}}^4 \sum_\balpha \qty[ \frac{2\pi^2}{3} \mathrm{B}_4\qty(\frac{\bphi\cdot\balpha}{2\pi}) - \tilde{a}^2 \mathrm{B}_2\qty(\frac{\bphi\cdot\balpha}{2\pi}) - \frac{11 \tilde{a}^4}{240\pi^2} ] \,.
  \end{split}
  \label{eq:Vg}
\end{equation}
The Polyakov-loop expectation value should be determined by a condition to minimize the above effective potential.
The difference from $V_{\text{opt}}(x)$ is the overall coefficient, and so the expectation value of the Polyakov loop is intact.  

\subsection{Singular screening with real acceleration effects}

\begin{figure}
    \centering
    \includegraphics[width=0.95\columnwidth]{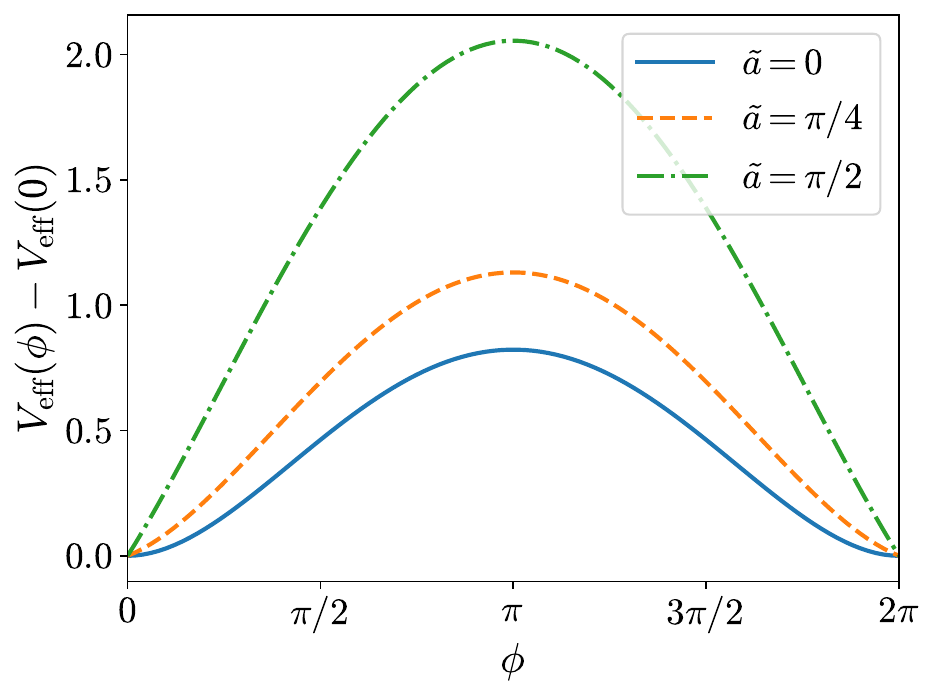}
    \caption{Physical Polyakov-loop effective potential for different real acceleration values as a function of $\phi$ for $SU(2)$ and $\tia$. The effective potential $V_{\text{eff}}$ is also made dimensionless by $T_{\text{loc}}^4$.}
    \label{fig:acc_opticalres}
\end{figure}

From the effective potential in Eq.~\eqref{eq:Vg}, we have found that the Polyakov-loop expectation value is not changed for any real acceleration.  This is because, as seen in Fig.~\ref{fig:acc_opticalres} for the simple $SU(2)$ case, the effective potential is never inverted, but the confining vacuum at $\expval{L}=0$ has an even larger energy with increasing $\tia$.

Hereafter, let us limit ourselves to the $SU(2)$ case for simplicity and analyze the properties of the effective potential around the perturbative vacuum as well as the confining vacuum.  Then, we can slightly rewrite the effective potential in Eq.~\eqref{eq:Vg} and analyze its properties around $\phi = \bphi \cdot \balpha = 0$ first.  That is, expanding the real-acceleration effective potential in Eq.~\eqref{eq:Vg} around
\(\phi=0\) leads to
\begin{equation}
V_{\text{eff}}(\tilde a)
=
V_{\text{eff}}(0)
+
2T_{\text{loc}}^4
\left[
\frac{\tilde a^2}{2\pi}|\phi|
+
\left(
\frac{1}{6}
-
\frac{\tilde a^2}{4\pi^2}
\right)\phi^2
+\cdots
\right].
\end{equation}
The nonanalytic \(|\phi|\) term originates from the \(\mathrm{B}_2\) contribution.
It turns \(\phi=0\) into a cusp minimum rather than a smooth quadratic
minimum.  Therefore the curvature at the minimum is not well-defined,
and the Debye mass extracted from this curvature becomes ill-defined in
the case of real acceleration.  This cusp becomes sharper as
\(\tilde a\) increases, reflecting the increasing slope discontinuity
at \(\phi=0\).

In contrast, we can safely expand the effective potential at another extremum at $\phi=\pi$ (i.e., the center-symmetric confining vacuum), resulting in
\begin{equation}
  \begin{split}
        & V_{\text{eff}}(x) \\
        &= T_{\text{loc}}^4  \qty[ \frac{(\phi-\pi)^4}{12\pi^2} - \qty(\inv{6}+\frac{\tilde{a}^2}{2\pi^2})(\phi-\pi)^2 + \dots
        ] \,,
  \end{split}
\end{equation}
where the ellipsis represents irrelevant terms that do not depend on $\phi$. 
The absence of a linear term implies that \(\phi=\pi\) is always an
extremum.  Since the coefficient of the quadratic term is negative,
this extremum is a local maximum.  Its instability is quantified by
\begin{equation}
\left.
\frac{\partial^2 V_{\text{eff}}}{\partial \phi^2}
\right|_{\phi=\pi}
=
-T_{\text{loc}}^4
\left(
\frac13+\frac{\tilde a^2}{\pi^2}
\right).
\end{equation}
The curvature is always negative, and
its magnitude increases with \(\tilde a\).  Therefore the
center-symmetric configuration becomes increasingly unstable as
\(\tilde a\) grows.  This is also reflected in the free-energy
difference
\begin{equation}
V_{\text{eff}}\bigr|_{\phi=\pi} - V_{\text{eff}}\bigr|_{\phi=0}
=
T_{\text{loc}}^4
\left(
\frac{\pi^2}{12}
+
\frac{\tilde a^2}{2}
\right)>0 \,,
\end{equation}
which increases with \(\tia\), indicating that the perturbative vacuum that is a center-broken extremum located at $\phi=0$ is further favored at larger \(\tia\).

\subsection{Analogy between imaginary acceleration and imaginary rotation effects}\label{sec:imaginary_acc}

\begin{figure}
    \centering
    \includegraphics[width=0.95\columnwidth]{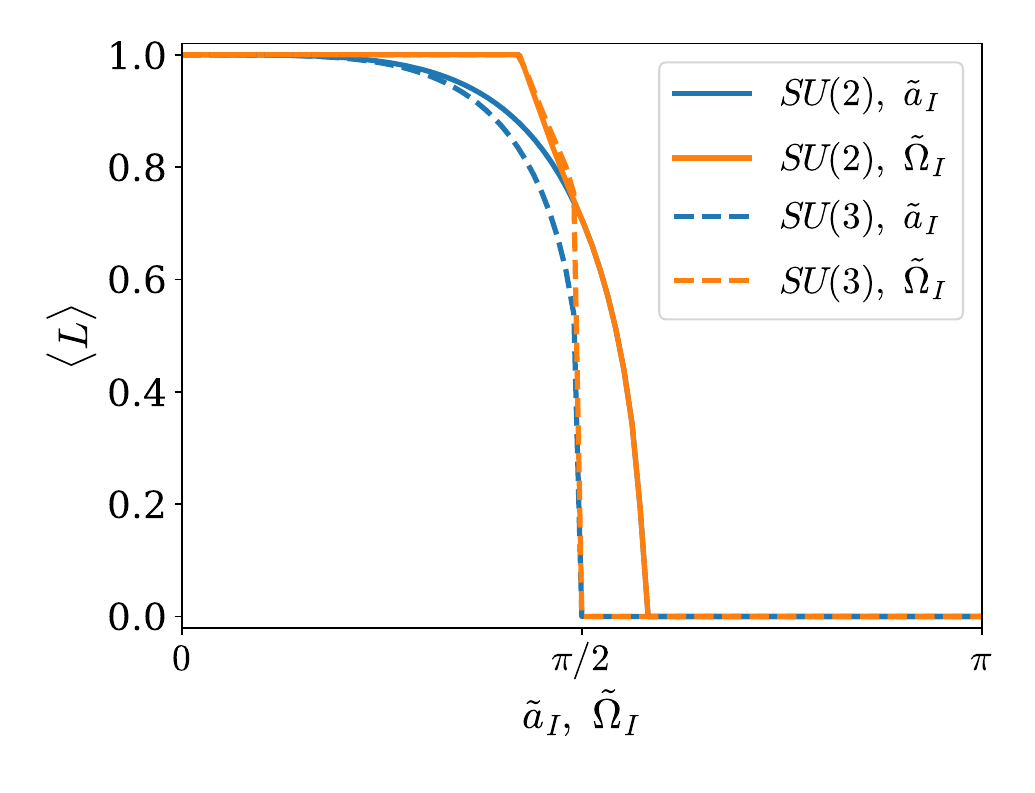}
    \caption{
    Polyakov loop $\expval{L}$ as a function of the imaginary
    acceleration $\tilde a_I$ and the imaginary angular velocity
    $\tilde\Omega_I$ for $SU(2)$ and $SU(3)$ gauge theories.
    }
    \label{fig:L_im}
\end{figure}

\begin{figure}
    \centering
    \includegraphics[width=0.95\columnwidth]{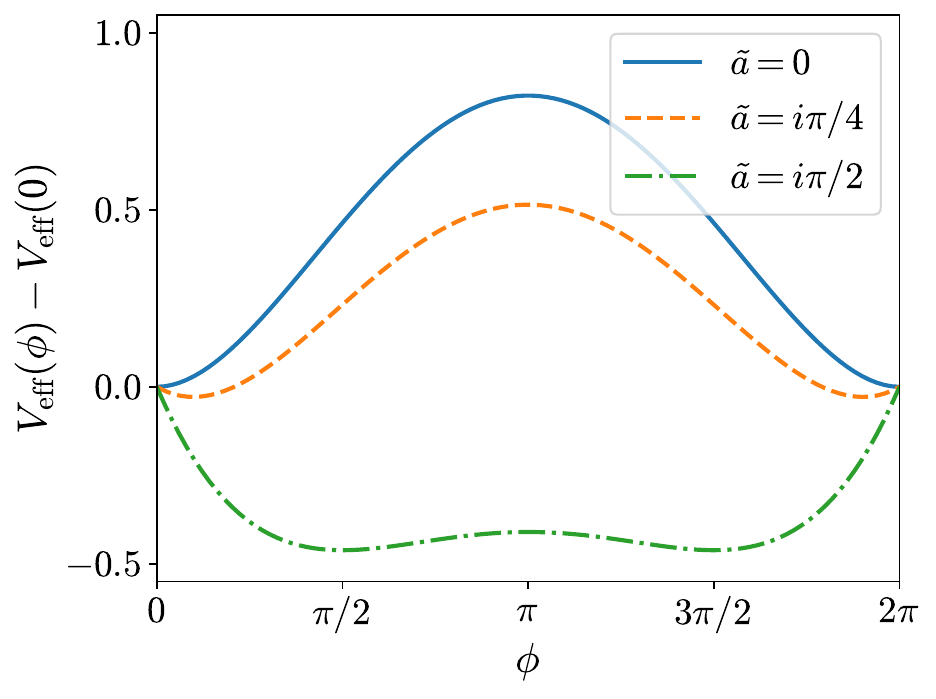}
    \caption{Polyakov-loop effective potential, Eq.~\eqref{eq:Veff_aI}, for different imaginary  acceleration values. $\phi$ is the background gauge parameter for $SU(2)$ and $\tilde{a}$ is dimensionless acceleration. The Polyakov-loop effective potential $V_\text{eff}$ is also made dimensionless by $T_{\text{loc}}^4$.}
    \label{fig:Veff_im}
\end{figure}

It is interesting to study how imaginary acceleration affects the confining
property of the system and to compare it with the imaginary-rotation case. 
By analytic continuation \(\tilde a\to -i\tilde a_I\), the
effective potential for $SU(2)$ Yang--Mills theory under imaginary acceleration
becomes
\begin{equation}
    \widetilde{V}_{\text{eff}}(x)
  = 2T_{\text{loc}}^4 
  \qty[ \frac{2\pi^2}{3} \mathrm{B}_4\qty(\frac{\phi}{2\pi}) + \tilde{a}_I^2 \mathrm{B}_2\qty(\frac{\phi}{2\pi}) + \dots ]\,.
  \label{eq:Veff_aI}
\end{equation}
Here, the overall factor \(2\) accounts for the two nonzero roots \(\alpha=\pm1\).
The mod-\(1\) prescription for \(\phi\) has been restored, and the
\(\phi\)-independent term has been omitted. Due to the flip of the sign of the \(\mathrm{B}_2\) term, imaginary acceleration tends
to drive the system toward the center-symmetric configuration, similar to
imaginary rotation. To make the comparison explicit, we recall that at the
rotation center the effective potential is given by~\cite{Chen:2022smf}
\begin{equation}
  V_\Omega\big|_{\tilde r=0} =
  \frac{2\pi^2 T_0^4}{3} \sum_{s=\pm 1}
  \mathrm{B}_4\left[ \left(
  \frac{\phi+s\tilde\Omega_I}{2\pi}
  \right)_{\mathrm{mod}\,1} \right] \,,
\label{eq:Veff_OmegaI}
\end{equation}
where \(\tilde\Omega_I=\beta_0\Omega_I\). At first sight,
Eqs.~\eqref{eq:Veff_aI} and~\eqref{eq:Veff_OmegaI} 
have rather different structures.  However, in the branch,
\begin{equation}
    \tilde\Omega_I < \phi < 2\pi-\tilde\Omega_I \,,
\end{equation}
the \(\phi\)-dependent part of the imaginary-rotation effective potential
coincides with that of the imaginary-acceleration effective potential, up to a
\(\phi\)-independent term, under the identification
\(\tilde\Omega_I\leftrightarrow \tilde a_I\).  For the \(SU(2)\)
minimum, this branch is relevant in the interval, $\frac{\pi}{2}<\tilde\Omega_I<\pi$.  For detailed branch analysis and comparison, see Appendix~\ref{app:branch}.

In Fig.~\ref{fig:L_im}, we show the Polyakov loop defined in
Eq.~\eqref{eq:L_def} under imaginary acceleration and imaginary rotation.
For completeness, we present the results for both \(SU(2)\) and \(SU(3)\)
Yang--Mills theories, although our analytic discussion is mainly based on
the \(SU(2)\) case.

Let us first discuss the \(SU(2)\) result.  In the region,
\begin{equation}
    0<\tilde\Omega_I \,,
    \qquad
    \tilde a_I<\frac{\pi}{2} \,,
\end{equation}
the imaginary-rotation and imaginary-acceleration curves are generally
different.  The imaginary-rotation result exhibits a plateau at small
\(\tilde\Omega_I\), whereas the imaginary-acceleration result decreases
smoothly with \(\tilde a_I\).  At
\begin{equation}
   \tilde\Omega_I=\tilde a_I=\frac{\pi}{2} \,,
\end{equation}
the two curves meet.  In the intermediate region
\begin{equation}
  \frac{\pi}{2}<\tilde\Omega_I \,,
  \qquad
  \tilde a_I<\frac{\pi}{\sqrt3} \,,
\end{equation}
the two curves coincide.  This is because, in this branch, the
\(\phi\)-dependent part of the imaginary-rotation effective potential takes
the same form as that of the imaginary-acceleration effective potential.
Both curves then reach
\(
\langle L\rangle=0
\)
at the critical value;
\begin{equation}
  \tilde a_{I,c}=\tilde\Omega_{I,c}=\frac{\pi}{\sqrt3} \,.
\end{equation}

The \(SU(3)\) result shows a qualitatively similar pattern.  In particular,
the imaginary-acceleration and imaginary-rotation curves also meet at
\(\tilde a_I=\tilde\Omega_I=\pi/2\).  However, in contrast to the \(SU(2)\)
case, where the Polyakov loop vanishes continuously at
\(\tilde a_{I,c}=\tilde\Omega_{I,c}=\pi/\sqrt{3}\), the \(SU(3)\) Polyakov
loop drops discontinuously to zero at
\(\tilde a_{I,c}=2\sqrt{5}\pi/9\) for imaginary acceleration and at \(\tilde\Omega_{I,c}=\pi/2\) for imaginary rotation, signaling a first-order phase transition.\footnote{We thank Rin Takada for pointing out the incorrect \(SU(3)\) critical value for imaginary acceleration in the original version of the manuscript.}

Another interesting feature of the imaginary-acceleration effective potential
is that its minimum is smooth in the \(SU(2)\) case, as shown in
Fig.~\ref{fig:Veff_im}.  This should be contrasted with the real-acceleration
case, where the nonanalytic term gives rise to a cusp at the minimum.
Therefore, the curvature of the imaginary-acceleration effective potential is
well-defined and can be used to characterize
the response of the Polyakov-loop background.  Its physical interpretation,
however, depends on the phase. In the deconfined phase,
the curvature of the effective potential can be identified with the
Debye screening mass squared,
\begin{equation}
  m_D^2(\tilde a_I) = \frac{g^2}{T_{\rm loc}^2}
  \left.
  \frac{\partial^2 V_{\text{eff}}}{\partial \phi^2}
  \right|_{\phi=\phi_{\mathrm{min}}} \,.
\end{equation}
For the imaginary-acceleration effective potential, this gives
\begin{equation}
  m_D^2(\tilde a_I) = \frac{2g^2T_{\text{loc}}^2}{3\pi^2}
  \left( \pi^2-3\tilde a_I^2 \right) \,,
  \quad
  0\leq \tilde a_I\leq \frac{\pi}{\sqrt3} \,.
\end{equation}
In the confined phase, the same curvature should no longer be interpreted
as an electric screening mass.  Instead, it corresponds to the string tension and behaves as
\begin{equation}
  \sigma \sim \frac{g^2T_{\text{loc}}^2}{3\pi^2}
  \left( 3\tilde a_I^2-\pi^2 \right) \,,
  \qquad
  \tilde a_I\geq \frac{\pi}{\sqrt3} \,.
\end{equation}

\section{Conclusion}

In this work, we investigated the Polyakov-loop effective potential of
accelerated gluonic matter.  We compared two complementary descriptions:
the direct formulation in Euclidean Rindler spacetime and the formulation
in the optical spacetime obtained by a conformal transformation.  Although
these two descriptions are geometrically related, we found that their
thermodynamic interpretations are not identical.  The local quantity
obtained from the Euclidean Rindler partition function is naturally related
to the spatial component of the energy-momentum tensor,
\(\langle T^z{}_z\rangle_{\rm ER}\), rather than directly to the physical
effective potential.  In contrast, the optical-spacetime calculation gives the
quantity satisfying the standard thermodynamic relation with the local
energy density.

The relation between the two descriptions can be understood through the
energy-momentum conservation law in Rindler spacetime.  In particular, for
the gluonic contribution considered here, one finds
\[
  \langle T^\tau{}_\tau\rangle_{\rm ER}
  =
  -3\langle T^z{}_z\rangle_{\rm ER},
\]
which leads to a thermodynamic relation between the optical effective
potential and the Rindler result.  This shows that the physical local
effective potential should be identified as
\[
  V_{\text{eff}}(x)
  =
  (az)^{-4} V_{\rm opt},
\]
up to terms independent of the Polyakov-loop background.  With this
identification, the optical calculation and the energy-momentum tensor in
the original Rindler spacetime give consistent physical results.

Using the physical effective potential, we found that real and imaginary
acceleration have opposite effects on the Polyakov-loop dynamics.  Real
acceleration favors the perturbative center-broken minimum and therefore
enhances the deconfined configuration, although the resulting nonanalyticity
makes the usual curvature definition of screening masses ill-defined.  By
contrast, imaginary acceleration favors the center-symmetric configuration,
in close analogy with imaginary rotation.  The analogy is, however, not
exact: imaginary rotation shifts the Polyakov-loop phase, while imaginary
acceleration changes the coefficient of the acceleration-induced
correction.  These results show that acceleration modifies not only the
Polyakov-loop effective potential itself but also the thermodynamic interpretation of
local quantities, making the distinction between the Rindler and optical
descriptions essential.

Some comments are in order.  In the optical calculation, we formulate the
gauge theory directly in the optical metric and impose the gauge-fixing
condition only after the conformal transformation.  This ordering is
essential.  In four dimensions the classical Yang--Mills action is
conformally invariant, so the physical gluonic sector can be consistently
defined in the optical spacetime.  By contrast, if one first fixes the gauge
in the Rindler metric and then performs the conformal transformation, the
gauge-fixing and ghost terms are transformed nontrivially.  As shown in
previous studies~\cite{Moretti:1996wd}, this latter procedure leads to an incorrect spectrum and
therefore to an incorrect thermodynamic potential.  Our calculation follows
the former prescription and is thus free from this ambiguity.

Second, the relation between the Rindler and optical descriptions found in
this work is not accidental.  It can be extended, for example, to the
conformal scalar field and massless fermion cases, and also holds in higher spacetime dimensions
\(D\).  In general, after subtracting temperature-independent vacuum contributions,
the corresponding relation for the thermal part takes the form
\begin{equation}
    (D-1)V_{\rm ER}
    =
    -
    \ue^{D\sigma}
    \partial_{\beta_0}
    \left(
        \beta_0 V_{\rm opt}
    \right) .
\end{equation}
This relation indicates that the quantity obtained directly from the
Rindler calculation is naturally associated with the energy-momentum-tensor component
\(\langle T^z{}_z\rangle\), whereas the optical calculation gives the
physical effective potential.  A more systematic discussion of this general relation
will be presented elsewhere.

Finally, a natural extension of the present work is to include fermionic
degrees of freedom.  In such an extension, massive fermions must be treated
explicitly.  This is a nontrivial problem because the fermion mass breaks
conformal invariance and may modify the simple thermodynamic relation
between the Rindler and optical descriptions.  It is therefore important to
clarify whether the identification of the physical local effective
potential continues to hold beyond the conformal limit.  We leave a
systematic study of massive fermionic contributions, including their impact
on the Polyakov-loop effective potential and possible implications for chiral dynamics
under acceleration, for future work.

\acknowledgments
The authors thank
Victor~Ambrus
for useful discussions.
This work was partially supported by JSPS KAKENHI [Grant Nos.\ 22H05118, 25K24464 and 26K00698 (K.F.) and JP26KJ0173 (Y.S.)].
H.L.C., Y.H.G., X.G.H., and Z.B.Z. are supported by the Natural Science Foundation of Shanghai (Grant No.\ 23JC1400200), the National Natural Science Foundation of China (Grants Nos.\ 12225502 and 12147101), and the National Key Research and Development Program of China (Grant No.\ 2022YFA1604900).

\appendix

\section{Derivation of Eq.~\eqref{eq:Tzz_ER}}
\label{app:der_Tzz}

\begin{figure}
    \centering
    \includegraphics[width=0.95\columnwidth]{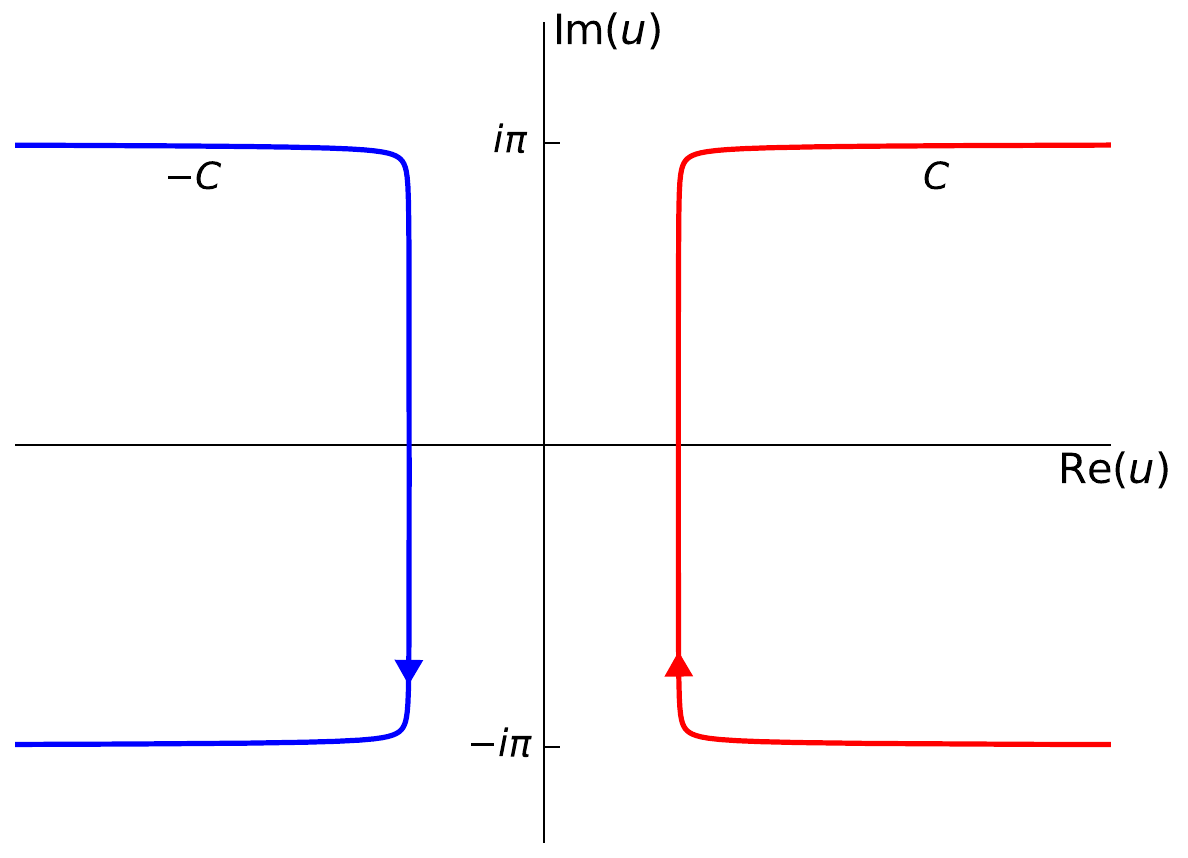}
    \caption{Contours $C$ and $-C$.}
    \label{fig:ContourC}
\end{figure}
\begin{figure}
    \centering
    \includegraphics[width=0.95\columnwidth]{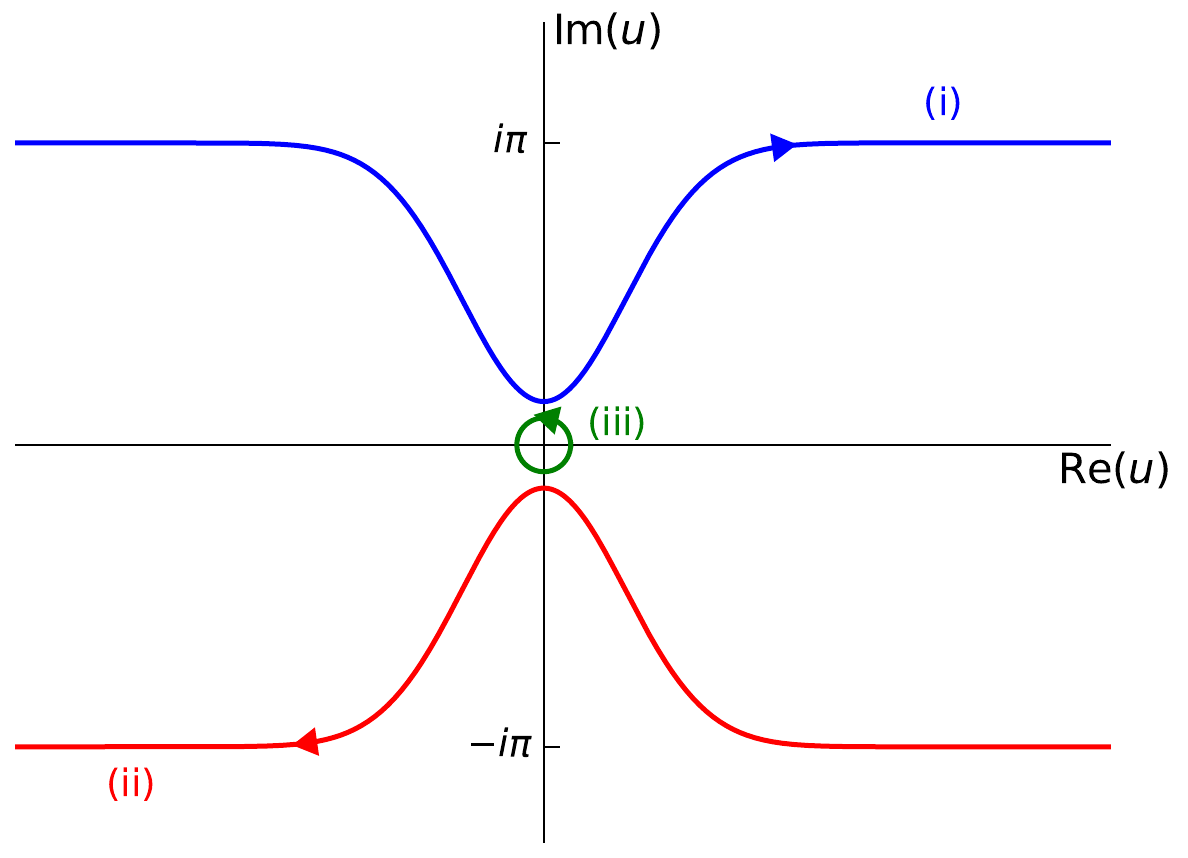}
    \caption{Contour $C_2$ after deformation.}
    \label{fig:ContourC2}
\end{figure}

We note that the modified Bessel function admits the following integral representation, with the integration contour $C$ shown in Fig.~\ref{fig:ContourC}:
\begin{equation}
    I_\nu(x) = \int_C  \frac{du}{2\pi i} \,  e^{x \cosh u - \nu u}\,.
\end{equation}
Since the Matsubara-frequency dependence appears only through \(\nu_n\) in Eq.~\eqref{eq:Tzz_integral}, we may first perform the summation explicitly as
\begin{equation}
  \sum_n e^{-|\nu_n |u}
  =
  \frac{e^{\bar\phi u}}{e^{v u}-1}
  +
  \frac{e^{v u}e^{-\bar\phi u}}{e^{v u}-1} \,.
\end{equation}
These two terms from the Matsubara sum can be combined by mapping the first one to the centrally reflected contour. Under the change of variable
\(u\to -u\), one finds,
\begin{equation}
  \int_C \frac{\ud u}{2\pi i}\,
  e^{\frac{zz'}{2s}\cosh u}
  \frac{e^{\bar\phi u}}{e^{v u}-1}
  =
  \int_{-C} \frac{\ud u}{2\pi i}\,
  e^{\frac{zz'}{2s}\cosh u}
  \frac{e^{v u}e^{-\bar\phi u}}{e^{v u}-1} \,,
\end{equation}
where $-C$ denotes the contour obtained from $C$ by the central reflection $u\to -u$; see Fig.~\ref{fig:ContourC}. Hence, defining
\begin{equation}
  C_2 = C\cup(-C) \,,
\end{equation}
we obtain
\begin{equation}
  \begin{split}
  &\sum_n K_{\nu_n}(s;z,z') \\
  &=\frac{1}{4s^2} e^{-\frac{z^2+z^{\prime2}}{4s}} \int_{C_2}\frac{\ud u}{2\pi i} \, e^{\frac{zz'}{2s}\cosh u} \, \frac{e^{v u} e^{-\bar\phi u}}{e^{v u}-1} \,.
  \end{split}
\end{equation}
The contour $C_2$ can be continuously deformed into three parts:
(i) a contour segment running from $i\pi-\infty$ to $i\pi+\infty$,
asymptotic to the horizontal line $\operatorname{Im}u=\pi$ at both ends,
(ii) a contour segment running from $-i\pi+\infty$ to $-i\pi-\infty$,
asymptotic to the horizontal line $\operatorname{Im}u=-\pi$ at both ends, 
and (iii) a small circle around $u = 0$.
Both (i) and (ii) cross the imaginary axis in the vicinity of the origin; see Fig.~\ref{fig:ContourC2} for the illustration of contour $C_2$ after deformation. Then, the kernel sum is decomposed as
\begin{equation}
  \begin{split}
  &\sum_n K_{\nu_n} (s;z,z') \\
  &= \frac{a\beta_0\,e^{-\frac{(z-z')^2}{4s}}}{8\pi s^2} + \frac{1}{4s^2} e^{-\frac{z^2+z^{\prime2}}{4s}} \int_{\chi} \frac{d u}{2\pi i}\, e^{\frac{zz'}{2s}\cosh u} \frac{e^{v u} e^{-\bar{\phi} u}}{e^{v u}-1} \,,
  \end{split}
\end{equation}
where $\chi$ denotes the joint contour of (i) and (ii).  The first term is divergent once the $s$ integration is performed; however, it clearly represents the Minkowski vacuum contribution, and can in principle be removed by an appropriate regularization.  Moreover, since it is independent of the background field, it does not affect the physical results of our interest.  After discarding the first term, it is quite straightforward to perform the $s$ integration to derive the following expression:
\begin{equation}
  \expval{T^{zz}}_{\text{ER}}
  = -\sum_{\balpha} \frac{1}{2\pi \tia} \int_{\chi}\frac{du}{2\pi i} \,\frac{1}{z^4(1-\cosh u)^2}\frac{e^{v u} e^{-\bar{\phi} u}}{e^{v u}-1} \,.
  \label{eq:Tzz_integ}
\end{equation}
Now we can close the contour at infinity and then evaluate the residue at $u=0$. Finally we obtain the expression in Eq.~\eqref{eq:Tzz_ER}.

\section{Relation between the surface term and the EMT trace}
\label{app:surface}

Let us discuss the discarded surface term in more detail.  Interestingly, one can show that this surface contribution is directly related to the EMT trace for a minimally coupled scalar field.  After performing the transverse momentum integral, the surface term takes the form of
\begin{equation}
  V_{\text{ER, surf}} = -\pi
  \sum_{\balpha, n} \int_0^\infty d\lambda
  \int_0^\infty \frac{d s}{s^2}\,
  e^{-s\lambda^2}
  \frac{\partial_z (z\Phi^*\partial_z\Phi)}{\lambda^2 z} \,.
\end{equation}
To see its relation to the EMT trace, we first construct the scalar propagator from
the eigenfunctions $\Phi$ in Eq.~\eqref{eq:Phi_mode}, i.e.,
\begin{equation}
  G_{\text{s}}(x,x') = \sum_n
  \int d^2\bk \int_0^\infty d\lambda \,
  \frac{\Phi(x)\Phi^*(x')}{\lambda^2+k^2} \,.
\end{equation}
For a minimally coupled scalar field, the EMT reads:
\begin{equation}
  \expval{T^\mu_{\;\;\nu}}_{\text{ms}} = \lim_{x'\to x}
  \qty( \partial^\mu\partial'_\nu - \frac{1}{2}\delta^\mu_{\;\;\nu} g^{\alpha\beta}\partial_\alpha\partial'_\beta )
  G_{\text{s}}(x,x') \,.
\end{equation}
Then, the EMT trace becomes
\begin{equation}
  \begin{split}
  \expval{T^\mu_{\;\;\mu}}_{\text{ms}} &= -\sum_n
  \int d^2\bk \int_0^\infty d\lambda \\
  &\qquad \times \qty[
  |\Phi|^2 + \frac{1}{\lambda^2+k^2} \frac{1}{z} \partial_z (z\Phi^*\partial_z\Phi) ] \,.
  \end{split}
\end{equation}
The first term is a contact term following from the completeness relation and
does not contribute after regularization.  The second term defines the renormalized trace denoted by $\expval{T^\mu_{\;\;\mu}}_{\text{ms,ren}}$.
After performing the transverse momentum integral and using integration by
parts in the proper-time variable, with the boundary term removed by the same
regularization, we obtain
\begin{equation}
  \expval{T^\mu_{\;\;\mu}}_{\text{ms,ren}} = \pi \sum_n
  \int_0^\infty d\lambda
  \int_0^\infty \frac{d s}{s^2}\,
  \frac{e^{-s\lambda^2}}{\lambda^2 z} \,
  \partial_z ( z\Phi^*\partial_z\Phi ) \,.
\end{equation}
Comparing this expression with the surface contribution, we finally find
\begin{equation}
  V_{\text{ER, surf}} = -\sum_{\balpha}
  \expval{T^\mu_{\;\;\mu}}_{\text{ms,ren}} \,.
\end{equation}

\section{Branch analysis of the effective potential with imaginary rotation}
\label{app:branch}

Let us derive the branchwise form of the effective potential with finite imaginary rotation discussed in Sec.~\ref{sec:imaginary_acc}.  We aim to explain why two results with finite imaginary rotation and imaginary acceleration coincide only in the interval \(\pi/2<\tilde\Omega_I<\pi/\sqrt{3}\), while they differ at smaller imaginary angular velocity.

\paragraph{Analysis for \(0\leq\phi(\tilde\Omega_I)\leq\tilde\Omega_I<\pi\)}:
In this branch, using the notation in the main text, we can write the imaginary-rotation effective potential as
\begin{equation}
  \begin{split}
    V_{\Omega}^{(1)}
    &= 2T_0^4
    \left[
    \frac{2\pi^2}{3}
    \mathrm{B}_4\left(\frac{\phi}{2\pi}\right)
    +
    \tilde\Omega_I^2
    \mathrm{B}_2\left(\frac{\phi}{2\pi}\right)
    \right] \\
    &\qquad + T_0^4 \left[
    \frac{(\phi-\tilde\Omega_I)^3}{3\pi}
    +
    \frac{\tilde\Omega_I^4}{12\pi^2} \right] \,.
  \end{split}
\end{equation}
The first term has the same structure as the imaginary-acceleration effective potential under the identification \(\tilde a_I\leftrightarrow\tilde\Omega_I\), whereas the second term is specific to the imaginary-rotation case.  The gap equation gives
\begin{equation}
  \frac{\partial V_{\Omega}^{(1)}}{\partial\phi} =
  \frac{T_0^4\phi}{3\pi^2}
  \left( \phi^2+3\tilde\Omega_I^2-6\pi\tilde\Omega_I+2\pi^2 \right) = 0 \,.
\end{equation}
This branch ends when the solution reaches the boundary at \(\phi=\tilde\Omega_I\). Substituting this boundary value for the gap equation gives the endpoint,
\begin{equation}
  \tilde\Omega_I = \frac{\pi}{2} \,.
\end{equation}
Therefore, this branch corresponds to the interval, $0\leq \tilde\Omega_I\leq \frac{\pi}{2}$.
\vspace{1em}

\paragraph{Analysis for \(\tilde\Omega_I<\phi(\tilde\Omega_I)\leq\pi\)}:
In this region, the \(\mathrm{mod}\,1\) operation does not affect the two
Bernoulli-polynomial arguments. Therefore, we can write
\begin{equation}
  V_{\Omega}^{(2)} = 2T_0^4
  \left[ \frac{2\pi^2}{3}
  \mathrm{B}_4 \left(\frac{\phi}{2\pi}\right) + \tilde\Omega_I^2
  \mathrm{B}_2\left(\frac{\phi}{2\pi}\right) \right] +
  T_0^4\frac{\tilde\Omega_I^4}{12\pi^2} \,.
\end{equation}
Thus, up to a \(\phi\)-independent term, this branch has exactly the same \(\phi\)-dependent structure as the imaginary-acceleration effective potential under \(\tilde a_I\leftrightarrow\tilde\Omega_I\). The gap equation is
\begin{equation}
  \frac{\partial V_{\Omega}^{(2)}}{\partial\phi} =
  \frac{T_0^4}{3\pi^2} (\phi-\pi)
  \left( \phi^2-2\pi\phi+3\tilde\Omega_I^2 \right) = 0 \,.
\end{equation}
The boundary between these branches is fixed by
\(\phi=\tilde\Omega_I\). Using the gap equation, this condition gives
\(\tilde\Omega_I=\pi/2\). Hence, this branch corresponds to $\frac{\pi}{2}<\tilde\Omega_I<\pi$.
Within this branch, the critical point is reached when the nontrivial minimum
merges with \(\phi=\pi\), which gives
\begin{equation}
  \tilde a_{I,c}=\tilde\Omega_{I,c}=\frac{\pi}{\sqrt3} \,.
\end{equation}
For \(\tilde\Omega_I>\pi/\sqrt3\), the nontrivial minimum disappears, and the
remaining minimum in this branch is \(\phi=\pi\).
\vspace{1em}

\paragraph{Analysis for \(\tilde\Omega_I>\pi\)}:
Using the periodicity of the Bernoulli
polynomial together with \(\mathrm{B}_4(1-x)=\mathrm{B}_4(x)\), one finds
\begin{equation}
  V_\Omega(\phi;\tilde\Omega_I) =
  V_\Omega(\phi;2\pi-\tilde\Omega_I) =
  V_\Omega(2\pi-\phi;\tilde\Omega_I) \,.
\end{equation}
Therefore, the region \(\tilde\Omega_I>\pi\) is related to
\(\tilde\Omega_I<\pi\) by the above symmetry. This periodic behavior originates from the
fact that \(\tilde\Omega_I\) acts as an imaginary chemical potential and shifts
the Polyakov-loop phase, \(\phi\rightarrow \phi\pm\tilde\Omega_I \).
This behavior is absent in the imaginary-acceleration case. There,
\(\tilde a_I\) does not shift the holonomic variable; instead, it appears only
as the coefficient of \(\mathrm{B}_2(\phi/2\pi)\). Consequently, the
imaginary-acceleration effective potential does not have an analogue of symmetry under \(\tilde\Omega_I\to2\pi-\tilde\Omega_I\). Thus, for large
\(\tilde a_I\), the minimum remains at \(\phi=\pi\),
rather than being periodically mapped back to the small-\(\tilde a_I\) region.

\bibliography{paper_Gluon_only}

@article{Zhong:2026ugc,
    author = "Zhong, Song-Ze and Deng, Xian-Gai and Huang, Xu-Guang and Ma, Yu-Gang",
    title = "{Fluid Acceleration in Heavy-Ion Collisions}",
    eprint = "2604.00302",
    archivePrefix = "arXiv",
    primaryClass = "nucl-th",
    journal = "",
    month = "3",
    year = "2026"
}

@article{Prokhorov:2025vak,
    author = "Prokhorov, G. Yu. and Shohonov, D. A. and Teryaev, O. V. and Tsegelnik, N. S. and Zakharov, V. I.",
    title = "{Modeling of acceleration in heavy-ion collisions: Occurrence of temperature below the Unruh temperature}",
    eprint = "2502.10146",
    archivePrefix = "arXiv",
    primaryClass = "nucl-th",
    doi = "10.1103/7m17-n41m",
    journal = "Phys. Rev. C",
    volume = "112",
    number = "6",
    pages = "064907",
    year = "2025"
}

@article{Moretti:1996ws,
    author = "Moretti, Valter",
    title = "{Euclidean thermal Green functions of photons in generalized Euclidean Rindler spaces for any Feynman - like gauge}",
    eprint = "hep-th/9607178",
    archivePrefix = "arXiv",
    reportNumber = "UTF-381",
    doi = "10.1142/S0217751X9700195X",
    journal = "Int. J. Mod. Phys. A",
    volume = "12",
    pages = "3787--3798",
    year = "1997"
}

@article{Lapygin:2025zhn,
    author = "Lapygin, Dmitry D. and Prokhorov, Georgy Yu. and Teryaev, Oleg V. and Zakharov, Valentin I.",
    title = "{Viscosity, entanglement, and acceleration}",
    eprint = "2502.18199",
    archivePrefix = "arXiv",
    primaryClass = "hep-th",
    doi = "10.1103/7r55-rqlz",
    journal = "Phys. Rev. D",
    volume = "112",
    number = "6",
    pages = "065012",
    year = "2025"
}

@article{Frolov:1987dz,
    author = "Frolov, Valeri P. and Serebryanyi, E. M.",
    title = "{Vacuum Polarization in the Gravitational Field of a Cosmic String}",
    doi = "10.1103/PhysRevD.35.3779",
    journal = "Phys. Rev. D",
    volume = "35",
    pages = "3779--3782",
    year = "1987"
}

@article{Page:1982fm,
    author = "Page, Don N.",
    title = "{Thermal Stress Tensors in Static Einstein Spaces}",
    reportNumber = "PRINT-82-0258 (PENN-STATE)",
    doi = "10.1103/PhysRevD.25.1499",
    journal = "Phys. Rev. D",
    volume = "25",
    pages = "1499",
    year = "1982"
}

@article{Weiss:1980rj,
    author = "Weiss, Nathan",
    title = "{The Effective Potential for the Order Parameter of Gauge Theories at Finite Temperature}",
    reportNumber = "UBC-81",
    doi = "10.1103/PhysRevD.24.475",
    journal = "Phys. Rev. D",
    volume = "24",
    pages = "475",
    year = "1981"
}

@article{Weiss:1981ev,
    author = "Weiss, Nathan",
    title = "{The Wilson Line in Finite Temperature Gauge Theories}",
    reportNumber = "Print-81-0743 (BRITISH COLUMBIA)",
    doi = "10.1103/PhysRevD.25.2667",
    journal = "Phys. Rev. D",
    volume = "25",
    pages = "2667",
    year = "1982"
}

@article{Gross:1980br,
    author = "Gross, David J. and Pisarski, Robert D. and Yaffe, Laurence G.",
    title = "{QCD and Instantons at Finite Temperature}",
    reportNumber = "PRINT-80-0538 (PRINCETON)",
    doi = "10.1103/RevModPhys.53.43",
    journal = "Rev. Mod. Phys.",
    volume = "53",
    pages = "43",
    year = "1981"
}

@article{Fukushima:2017csk,
    author = "Fukushima, Kenji and Skokov, Vladimir",
    title = "{Polyakov loop modeling for hot QCD}",
    eprint = "1705.00718",
    archivePrefix = "arXiv",
    primaryClass = "hep-ph",
    doi = "10.1016/j.ppnp.2017.05.002",
    journal = "Prog. Part. Nucl. Phys.",
    volume = "96",
    pages = "154--199",
    year = "2017"
}

@article{Chen:2022smf,
    author = "Chen, Shi and Fukushima, Kenji and Shimada, Yusuke",
    title = "{Perturbative Confinement in Thermal Yang-Mills Theories Induced by Imaginary Angular Velocity}",
    eprint = "2207.12665",
    archivePrefix = "arXiv",
    primaryClass = "hep-ph",
    doi = "10.1103/PhysRevLett.129.242002",
    journal = "Phys. Rev. Lett.",
    volume = "129",
    number = "24",
    pages = "242002",
    year = "2022"
}

@article{Unruh:1976db,
    author = "Unruh, W. G.",
    title = "{Notes on black hole evaporation}",
    doi = "10.1103/PhysRevD.14.870",
    journal = "Phys. Rev. D",
    volume = "14",
    pages = "870",
    year = "1976"
}

@article{Crispino:2007eb,
    author = "Crispino, Luis C. B. and Higuchi, Atsushi and Matsas, George E. A.",
    title = "{The Unruh effect and its applications}",
    eprint = "0710.5373",
    archivePrefix = "arXiv",
    primaryClass = "gr-qc",
    doi = "10.1103/RevModPhys.80.787",
    journal = "Rev. Mod. Phys.",
    volume = "80",
    pages = "787--838",
    year = "2008"
}

@article{Fulling:1972md,
    author = "Fulling, Stephen A.",
    title = "{Nonuniqueness of canonical field quantization in Riemannian space-time}",
    doi = "10.1103/PhysRevD.7.2850",
    journal = "Phys. Rev. D",
    volume = "7",
    pages = "2850--2862",
    year = "1973"
}

@article{Davies:1974th,
    author = "Davies, P. C. W.",
    title = "{Scalar particle production in Schwarzschild and Rindler metrics}",
    doi = "10.1088/0305-4470/8/4/022",
    journal = "J. Phys. A",
    volume = "8",
    pages = "609--616",
    year = "1975"
}

@article{Unruh:1983ac,
    author = "Unruh, William G. and Weiss, Nathan",
    title = "{Acceleration Radiation in Interacting Field Theories}",
    reportNumber = "Print-83-1008 (BRITISH COLUMBIA)",
    doi = "10.1103/PhysRevD.29.1656",
    journal = "Phys. Rev. D",
    volume = "29",
    pages = "1656",
    year = "1984"
}

@article{Kharzeev:2005iz,
    author = "Kharzeev, Dmitri and Tuchin, Kirill",
    title = "{From color glass condensate to quark gluon plasma through the event horizon}",
    eprint = "hep-ph/0501234",
    archivePrefix = "arXiv",
    reportNumber = "BNL-NT-05-2",
    doi = "10.1016/j.nuclphysa.2005.03.001",
    journal = "Nucl. Phys. A",
    volume = "753",
    pages = "316--334",
    year = "2005"
}

@article{Ohsaku:2004rv,
    author = "Ohsaku, Tadafumi",
    title = "{Dynamical chiral symmetry breaking and its restoration for an accelerated observer}",
    eprint = "hep-th/0407067",
    archivePrefix = "arXiv",
    doi = "10.1016/j.physletb.2004.08.019",
    journal = "Phys. Lett. B",
    volume = "599",
    pages = "102--110",
    year = "2004"
}

@article{Ebert:2006bh,
    author = "Ebert, D. and Zhukovsky, V. Ch.",
    title = "{Restoration of Dynamically Broken Chiral and Color Symmetries for an Accelerated Observer}",
    eprint = "hep-th/0612009",
    archivePrefix = "arXiv",
    reportNumber = "HU-EP-06-45",
    doi = "10.1016/j.physletb.2006.12.013",
    journal = "Phys. Lett. B",
    volume = "645",
    pages = "267--274",
    year = "2007"
}

@article{Peeters:2007ti,
    author = "Peeters, Kasper and Zamaklar, Marija",
    title = "{Dissociation by acceleration}",
    eprint = "0711.3446",
    archivePrefix = "arXiv",
    primaryClass = "hep-th",
    reportNumber = "ITP-UU-07-59, SPIN-07-45, DCPT-07-63",
    doi = "10.1088/1126-6708/2008/01/038",
    journal = "JHEP",
    volume = "01",
    pages = "038",
    year = "2008"
}

@article{Paredes:2008cr,
    author = "Paredes, Angel and Peeters, Kasper and Zamaklar, Marija",
    title = "{Temperature versus acceleration: The Unruh effect for holographic models}",
    eprint = "0812.0981",
    archivePrefix = "arXiv",
    primaryClass = "hep-th",
    reportNumber = "ITP-UU-08-71, SPIN-08-54, DCPT-08-65",
    doi = "10.1088/1126-6708/2009/04/015",
    journal = "JHEP",
    volume = "04",
    pages = "015",
    year = "2009"
}

@article{Ghoroku:2010sp,
    author = "Ghoroku, Kazuo and Ishihara, Masafumi and Kubo, Kouki and Taminato, Tomoki",
    title = "{Accelerated Quark and Holography for Confining Gauge theory}",
    eprint = "1010.4396",
    archivePrefix = "arXiv",
    primaryClass = "hep-th",
    reportNumber = "FIT-HE--10-02",
    doi = "10.1103/PhysRevD.83.024020",
    journal = "Phys. Rev. D",
    volume = "83",
    pages = "024020",
    year = "2011"
}

@article{Castorina:2012yg,
    author = "Castorina, P. and Finocchiaro, M.",
    title = "{Symmetry Restoration By Acceleration}",
    eprint = "1207.3677",
    archivePrefix = "arXiv",
    primaryClass = "hep-th",
    doi = "10.4236/jmp.2012.311209",
    journal = "J. Mod. Phys.",
    volume = "3",
    pages = "1703",
    year = "2012"
}

@article{Takeuchi:2015nga,
    author = "Takeuchi, Shingo",
    title = "{Bose{\textendash}Einstein condensation in the Rindler space}",
    eprint = "1501.07471",
    archivePrefix = "arXiv",
    primaryClass = "hep-th",
    doi = "10.1016/j.physletb.2015.09.013",
    journal = "Phys. Lett. B",
    volume = "750",
    pages = "209--217",
    year = "2015"
}

@inproceedings{Dobado:2017dvs,
    author = "Dobado, Antonio",
    title = "{Spontaneous symmetry breaking and the Unruh effect}",
    booktitle = "{4th International Conference on the Nature and Ontology of Spacetime}",
    eprint = "1703.05675",
    archivePrefix = "arXiv",
    primaryClass = "gr-qc",
    pages = "161--173",
    month = "3",
    year = "2017"
}

@article{Dobado:2017xxb,
    author = "Dobado, Antonio",
    title = "{Brout-Englert-Higgs mechanism for accelerating observers}",
    eprint = "1710.01564",
    archivePrefix = "arXiv",
    primaryClass = "gr-qc",
    doi = "10.1103/PhysRevD.96.085009",
    journal = "Phys. Rev. D",
    volume = "96",
    number = "8",
    pages = "085009",
    year = "2017"
}

@inproceedings{Casado-Turrion:2019xta,
    author = "Casado-Turri{\'o}n, Adri{\'a}n and Dobado, Antonio",
    title = "{Chiral symmetry breaking and the Unruh effect}",
    booktitle = "{15th Marcel Grossmann Meeting on Recent Developments in Theoretical and Experimental General Relativity, Astrophysics, and Relativistic Field Theories}",
    eprint = "1901.06964",
    archivePrefix = "arXiv",
    primaryClass = "hep-th",
    doi = "10.1142/9789811258251_0206",
    month = "1",
    year = "2019"
}

@article{Casado-Turrion:2019gbg,
    author = "Casado-Turri{\'o}n, Adri{\'a}n and Dobado, Antonio",
    title = "{Triggering the QCD phase transition through the Unruh effect: chiral symmetry restoration for uniformly accelerated observers}",
    eprint = "1905.11179",
    archivePrefix = "arXiv",
    primaryClass = "hep-ph",
    doi = "10.1103/PhysRevD.99.125018",
    journal = "Phys. Rev. D",
    volume = "99",
    number = "12",
    pages = "125018",
    year = "2019"
}

@article{Kou:2024dml,
    author = "Kou, Wei and Chen, Xurong",
    title = "{Locating quark-antiquark string breaking in QCD through chiral symmetry restoration and Hawking-Unruh effect}",
    eprint = "2405.18697",
    archivePrefix = "arXiv",
    primaryClass = "hep-ph",
    doi = "10.1016/j.physletb.2024.138942",
    journal = "Phys. Lett. B",
    volume = "856",
    pages = "138942",
    year = "2024"
}

@article{Benic:2015qha,
    author = "Benic, Sanjin and Fukushima, Kenji",
    title = "{Unruh effect and condensate in and out of an accelerated vacuum}",
    eprint = "1503.05790",
    archivePrefix = "arXiv",
    primaryClass = "hep-th",
    journal = "",
    month = "3",
    year = "2015"
}

@article{Salluce:2024jlj,
    author = "Salluce, Domenico Giuseppe and Pasini, Marco and Flachi, Antonino and Pittelli, Antonio and Ansoldi, Stefano",
    title = "{Symmetry restoration and uniformly accelerated observers in Minkowski spacetime}",
    eprint = "2401.16483",
    archivePrefix = "arXiv",
    primaryClass = "hep-th",
    doi = "10.1007/JHEP05(2024)218",
    journal = "JHEP",
    volume = "05",
    pages = "218",
    year = "2024"
}

@article{Zhu:2025pxh,
    author = "Zhu, Zhi-Bin and Chen, Hao-Lei and Huang, Xu-Guang",
    title = "{Chiral symmetry breaking in accelerating and rotating frames}",
    eprint = "2511.03230",
    archivePrefix = "arXiv",
    primaryClass = "hep-ph",
    doi = "10.1103/yxj9-33z6",
    journal = "Phys. Rev. D",
    volume = "113",
    number = "3",
    pages = "034005",
    year = "2026"
}

@article{Chernodub:2025ovo,
    author = "Chernodub, Maxim N.",
    title = "{Acceleration as refrigeration: Acceleration-induced spontaneous symmetry breaking in thermal medium}",
    eprint = "2501.16129",
    archivePrefix = "arXiv",
    primaryClass = "hep-th",
    journal = "",
    month = "1",
    year = "2025"
}

@article{Chernodub:2024wis,
    author = "Chernodub, M. N. and Goy, V. A. and Molochkov, A. V. and Stepanov, D. V. and Pochinok, A. S.",
    title = "{Extreme Softening of QCD Phase Transition under Weak Acceleration: First-Principles Monte~Carlo Results for Gluon Plasma}",
    eprint = "2409.01847",
    archivePrefix = "arXiv",
    primaryClass = "hep-lat",
    doi = "10.1103/PhysRevLett.134.111904",
    journal = "Phys. Rev. Lett.",
    volume = "134",
    number = "11",
    pages = "111904",
    year = "2025"
}

@article{Braguta:2026nfy,
    author = "Braguta, Victor V. and Goy, Vladimir A. and Dey, Jayanta and Roenko, Artem A.",
    title = "{Spatial confinement-deconfinement transition in accelerated gluodynamics within lattice simulation}",
    eprint = "2602.20970",
    archivePrefix = "arXiv",
    primaryClass = "hep-lat",
    month = "2",
    year = "2026",
    journal = ""
}

@article{Dowker:1994fi,
    author = "Dowker, J. S.",
    title = "{Remarks on geometric entropy}",
    eprint = "hep-th/9401159",
    archivePrefix = "arXiv",
    reportNumber = "MUTP-94-2",
    doi = "10.1088/0264-9381/11/4/001",
    journal = "Class. Quant. Grav.",
    volume = "11",
    pages = "L55--L60",
    year = "1994"
}

@article{Emparan:1994qa,
    author = "Emparan, R.",
    title = "{Heat kernels and thermodynamics in Rindler space}",
    eprint = "hep-th/9407064",
    archivePrefix = "arXiv",
    reportNumber = "EHU-FT-94-5",
    doi = "10.1103/PhysRevD.51.5716",
    journal = "Phys. Rev. D",
    volume = "51",
    pages = "5716--5719",
    year = "1995"
}

@article{Fursaev:1995ef,
    author = "Fursaev, Dmitri V. and Solodukhin, Sergey N.",
    title = "{On the description of the Riemannian geometry in the presence of conical defects}",
    eprint = "hep-th/9501127",
    archivePrefix = "arXiv",
    reportNumber = "JINR-E2-95-28, JINR, E2-95-28",
    doi = "10.1103/PhysRevD.52.2133",
    journal = "Phys. Rev. D",
    volume = "52",
    pages = "2133--2143",
    year = "1995"
}

@article{Fursaev:1996uz,
    author = "Fursaev, Dmitri V. and Miele, Gennaro",
    title = "{Cones, spins and heat kernels}",
    eprint = "hep-th/9605153",
    archivePrefix = "arXiv",
    reportNumber = "ALBERTA-THY-17-96, DSF-T-23-96",
    doi = "10.1016/S0550-3213(96)00631-1",
    journal = "Nucl. Phys. B",
    volume = "484",
    pages = "697--723",
    year = "1997"
}

@article{Kabat:1995eq,
    author = "Kabat, Daniel N.",
    title = "{Black hole entropy and entropy of entanglement}",
    eprint = "hep-th/9503016",
    archivePrefix = "arXiv",
    reportNumber = "RU-95-06",
    doi = "10.1016/0550-3213(95)00443-V",
    journal = "Nucl. Phys. B",
    volume = "453",
    pages = "281--299",
    year = "1995"
}

@inproceedings{Iellici:1996jx,
    author = "Iellici, D. and Moretti, V.",
    title = "{Kabat's surface terms in the zeta function approach}",
    booktitle = "{12th Italian Conference on General Relativity and Gravitational Physics}",
    eprint = "hep-th/9703088",
    archivePrefix = "arXiv",
    pages = "317--321",
    month = "9",
    year = "1996"
}

@article{Moretti:1996wd,
    author = "Moretti, Valter and Iellici, Devis",
    title = "{Optical approach for the thermal partition function of photons}",
    eprint = "hep-th/9610180",
    archivePrefix = "arXiv",
    reportNumber = "UTF-388",
    doi = "10.1103/PhysRevD.55.3552",
    journal = "Phys. Rev. D",
    volume = "55",
    pages = "3552--3563",
    year = "1997"
}

@article{Linet:1987vz,
    author = "Linet, B.",
    title = "{Quantum Field Theory in the Space-time of a Cosmic String}",
    doi = "10.1103/PhysRevD.35.536",
    journal = "Phys. Rev. D",
    volume = "35",
    pages = "536--539",
    year = "1987"
}

@article{Fursaev:1993qk,
    author = "Fursaev, D. V.",
    title = "{The Heat kernel expansion on a cone and quantum fields near cosmic strings}",
    eprint = "hep-th/9309050",
    archivePrefix = "arXiv",
    reportNumber = "JINR-E2-93-291",
    doi = "10.1088/0264-9381/11/6/008",
    journal = "Class. Quant. Grav.",
    volume = "11",
    pages = "1431--1444",
    year = "1994"
}

@article{Gusev:1998rp,
    author = "Gusev, Yu. V. and Zelnikov, A. I",
    title = "{Finite temperature nonlocal effective action for quantum fields in curved space}",
    eprint = "hep-th/9807038",
    archivePrefix = "arXiv",
    reportNumber = "ALBERTA-THY-11-98",
    doi = "10.1103/PhysRevD.59.024002",
    journal = "Phys. Rev. D",
    volume = "59",
    pages = "024002",
    year = "1999"
}

@article{Gibbons:1976pt,
    author = "Gibbons, G. W. and Perry, M. J.",
    editor = "Gibbons, G. W. and Hawking, S. W.",
    title = "{Black Holes and Thermal Green's Functions}",
    reportNumber = "PRINT-76-0846 (CAMBRIDGE)",
    doi = "10.1098/rspa.1978.0022",
    journal = "Proc. Roy. Soc. Lond. A",
    volume = "358",
    pages = "467--494",
    year = "1978"
}

@article{Dowker:1978md,
    author = "Dowker, J. S. and Kennedy, Gerard",
    title = "{Finite Temperature and Boundary Effects in Static Space-Times}",
    reportNumber = "PRINT-78-0112 (MANCHESTER)",
    doi = "10.1088/0305-4470/11/5/020",
    journal = "J. Phys. A",
    volume = "11",
    pages = "895",
    year = "1978"
}

@article{Bytsenko:1995ym,
    author = "Bytsenko, Andrei A. and Cognola, Guido and Zerbini, Sergio",
    title = "{Finite temperature effects for massive fields in d-dimensional Rindler - like spaces}",
    eprint = "hep-th/9508104",
    archivePrefix = "arXiv",
    reportNumber = "UTF-357",
    doi = "10.1016/0550-3213(95)00585-4",
    journal = "Nucl. Phys. B",
    volume = "458",
    pages = "267--290",
    year = "1996"
}

@article{Cognola:1997dv,
    author = "Cognola, Guido",
    title = "{Thermodynamics of scalar fields in Kerr's geometry}",
    eprint = "gr-qc/9710118",
    archivePrefix = "arXiv",
    reportNumber = "UTF-407",
    doi = "10.1103/PhysRevD.57.6292",
    journal = "Phys. Rev. D",
    volume = "57",
    pages = "6292--6296",
    year = "1998"
}

@article{Dowker:1989gp,
    author = "Dowker, J. S. and Schofield, John P.",
    title = "{Chemical Potentials in Curved Space}",
    reportNumber = "MUTP-25/88",
    doi = "10.1016/0550-3213(89)90295-2",
    journal = "Nucl. Phys. B",
    volume = "327",
    pages = "267--284",
    year = "1989"
}

@article{Sonego:2010vy,
    author = "Sonego, Sebastiano",
    title = "{Ultrastatic spacetimes}",
    eprint = "1004.1714",
    archivePrefix = "arXiv",
    primaryClass = "gr-qc",
    doi = "10.1063/1.3485599",
    journal = "J. Math. Phys.",
    volume = "51",
    pages = "092502",
    year = "2010"
}

@article{Frolov:1998vs,
    author = "Frolov, Valeri P. and Fursaev, D. V.",
    title = "{Thermal fields, entropy, and black holes}",
    eprint = "hep-th/9802010",
    archivePrefix = "arXiv",
    reportNumber = "ALBERTA-THY-03-98",
    doi = "10.1088/0264-9381/15/8/001",
    journal = "Class. Quant. Grav.",
    volume = "15",
    pages = "2041--2074",
    year = "1998"
}

@article{Vassilevich:2003xt,
    author = "Vassilevich, D. V.",
    title = "{Heat kernel expansion: User's manual}",
    eprint = "hep-th/0306138",
    archivePrefix = "arXiv",
    doi = "10.1016/j.physrep.2003.09.002",
    journal = "Phys. Rept.",
    volume = "388",
    pages = "279--360",
    year = "2003"
}

@article{Solodukhin:1994yz,
    author = "Solodukhin, Sergei N.",
    title = "{The Conical singularity and quantum corrections to entropy of black hole}",
    eprint = "hep-th/9407001",
    archivePrefix = "arXiv",
    reportNumber = "JINR-E2-94-246",
    doi = "10.1103/PhysRevD.51.609",
    journal = "Phys. Rev. D",
    volume = "51",
    pages = "609--617",
    year = "1995"
}

@article{Fursaev:1997th,
    author = "Fursaev, Dmitri V.",
    title = "{Euclidean and canonical formulations of statistical mechanics in the presence of killing horizons}",
    eprint = "hep-th/9709213",
    archivePrefix = "arXiv",
    reportNumber = "ALBERTA-THY-20-97",
    doi = "10.1016/S0550-3213(98)00197-7",
    journal = "Nucl. Phys. B",
    volume = "524",
    pages = "447--468",
    year = "1998"
}

@article{Diakonov:2023jdk,
    author = "Diakonov, Dmitrii",
    title = "{Is the Euclidean path integral always equal to the thermal partition function?}",
    eprint = "2310.08522",
    archivePrefix = "arXiv",
    primaryClass = "hep-th",
    doi = "10.1007/JHEP04(2024)077",
    journal = "JHEP",
    volume = "04",
    pages = "077",
    year = "2024"
}

@article{Becattini:2016stj,
    author = "Becattini, F.",
    title = "{Thermodynamic equilibrium in relativity: four-temperature, Killing vectors and Lie derivatives}",
    eprint = "1606.06605",
    archivePrefix = "arXiv",
    primaryClass = "gr-qc",
    doi = "10.5506/APhysPolB.47.1819",
    journal = "Acta Phys. Polon. B",
    volume = "47",
    pages = "1819",
    year = "2016"
}

@article{Zubarev:1979afm,
    author = "Zubarev, D. N. and Prozorkevich, A. V. and Smolyanskii, S. A.",
    title = "{Derivation of nonlinear generalized equations of quantum relativistic hydrodynamics}",
    doi = "10.1007/BF01032069",
    journal = "Theor. Math. Phys.",
    volume = "40",
    number = "3",
    pages = "821--831",
    year = "1979"
}

@article{Prokhorov:2019cik,
    author = "Prokhorov, George Y. and Teryaev, Oleg V. and Zakharov, Valentin I.",
    title = "{Unruh effect for fermions from the Zubarev density operator}",
    eprint = "1903.09697",
    archivePrefix = "arXiv",
    primaryClass = "hep-th",
    doi = "10.1103/PhysRevD.99.071901",
    journal = "Phys. Rev. D",
    volume = "99",
    number = "7",
    pages = "071901",
    year = "2019"
}

@article{Prokhorov:2019yft,
    author = "Prokhorov, Georgy Y. and Teryaev, Oleg V. and Zakharov, Valentin I.",
    title = "{Unruh effect universality: emergent conical geometry from density operator}",
    eprint = "1911.04545",
    archivePrefix = "arXiv",
    primaryClass = "hep-th",
    doi = "10.1007/JHEP03(2020)137",
    journal = "JHEP",
    volume = "03",
    pages = "137",
    year = "2020"
}

@article{Becattini:2020qol,
    author = "Becattini, F. and Buzzegoli, M. and Palermo, A.",
    title = "{Exact equilibrium distributions in statistical quantum field theory with rotation and acceleration: scalar field}",
    eprint = "2007.08249",
    archivePrefix = "arXiv",
    primaryClass = "hep-th",
    doi = "10.1007/JHEP02(2021)101",
    journal = "JHEP",
    volume = "02",
    pages = "101",
    year = "2021"
}

@article{Palermo:2021hlf,
    author = "Palermo, Andrea and Buzzegoli, Matteo and Becattini, Francesco",
    title = "{Exact equilibrium distributions in statistical quantum field theory with rotation and acceleration: Dirac field}",
    eprint = "2106.08340",
    archivePrefix = "arXiv",
    primaryClass = "hep-th",
    doi = "10.1007/JHEP10(2021)077",
    journal = "JHEP",
    volume = "10",
    pages = "077",
    year = "2021"
}

@phdthesis{Palermo:2023ews,
    author = "Palermo, Andrea",
    title = "{Spin polarization in the strongly interacting QCD matter at global and local equilibrium}",
    school = "U. Florence (main), Goethe U., Frankfurt (main)",
    year = "2023"
}

@article{Becattini:2025oyi,
    author = "Becattini, Francesco and Singh, Rajeev",
    title = "{On the local thermodynamic relations in relativistic spin hydrodynamics}",
    eprint = "2506.20681",
    archivePrefix = "arXiv",
    primaryClass = "nucl-th",
    doi = "10.1140/epjc/s10052-025-15071-3",
    journal = "Eur. Phys. J. C",
    volume = "85",
    number = "11",
    pages = "1338",
    year = "2025"
}

@article{Ambrus:2025dca,
    author = "Ambrus, Victor E. and Geci{\'c}, Aleksandar",
    title = "{Thermodynamics of rotating fermions}",
    eprint = "2509.17640",
    archivePrefix = "arXiv",
    primaryClass = "hep-th",
    journal = "",
    month = "9",
    year = "2025"
}
\bibliographystyle{apsrev4-2}
\end{document}